\documentclass[
  aps,
  pra,
  reprint,
  longbibliography,
  amsmath,
  amssymb,
  floatfix
]{revtex4-2}

\usepackage{amsfonts,amssymb,amsmath}
\usepackage[]{graphics,graphicx,epsfig}
\usepackage{amssymb}
\usepackage{multirow}
\usepackage{comment}
\usepackage{epstopdf}
\usepackage{graphicx}
\usepackage{natbib}
\usepackage{color}

\newcommand{\ket}[1]{|{#1}\rangle}
\newcommand{\bra}[1]{\langle{#1}|}
\usepackage{array}

\newcommand{\R}{\mathbb{R}}

\newcommand{\KG}{K_{\mathrm G}}

\newcommand{\sgn}{\operatorname{sign}}

\newtheorem{conjecture}{Conjecture}

\usepackage[
  colorlinks=true,
  citecolor=blue,
  linkcolor=blue,
  urlcolor=blue
]{hyperref}

\begin{document}

\title{Heuristic lower bounds on real Grothendieck constants of finite and infinite order}

\author{Erika Bene}
\affiliation{HUN-REN Institute for Nuclear Research, PO Box 51, H-4001 Debrecen, Hungary}

\author{Tamás Vértesi}
\affiliation{HUN-REN Institute for Nuclear Research, PO Box 51, H-4001 Debrecen, Hungary}

\begin{abstract}
We propose candidate lower bounds on the real Grothendieck constants $\KG(d)$
and their infinite-dimensional limit $\KG$ using rotationally invariant cubic
kernels. For the continuous construction, we identify an explicit threshold
above which hemispherical binary strategies are unstable under cubic boundary
deformations. Supported by numerical evidence, we conjecture that
hemispherical strategies are globally optimal at this threshold. This
conjecture would imply $\KG\geq9\pi/16\simeq1.767146$, improving the rigorous
lower bound $6\pi/11\simeq1.713596$. We also construct finite symmetric
coefficient matrices from unit vectors in dimensions up to $24$. Large-scale
see-saw optimization supports the continuous predictions and suggests improved
lower bounds on $\KG(d)$ in the studied dimensions from $4$ to $24$,
conditional on the conjectured binary optima being exact. Selected matrices
also yield finite-setting candidate dimension witnesses when interpreted as
bipartite correlation Bell functionals. If the corresponding binary upper
bounds hold, these witnesses would certify a local Hilbert-space dimension of at
least five for each party.
\end{abstract}

\maketitle

\section{Introduction}
\label{sec:intro}

The Grothendieck inequality was originally formulated in the language of
tensor products of Banach spaces \cite{Grothendieck1953,Pisier2012}. It now
connects functional analysis, combinatorial optimization, and theoretical
computer science~\cite{AlonNaor2006}. Its relevance to quantum information
follows from Tsirelson's characterization of bipartite quantum correlations
\cite{Tsirelson1980,Tsirelson1987}. In particular, the real Grothendieck
constant bounds the quantum-to-classical ratio of correlation Bell
expressions, or equivalently of biases in two-player XOR games
\cite{Tsirelson1993,Acin2006}.

In its real matrix form, the inequality compares optimization over binary
signs with optimization over unit vectors. Let
$M=(M_{xy})\in\mathbb{R}^{m\times m}$. For an integer $d\geq1$, define
\begin{equation}
  Q_d(M)
  :=
  \max
  \sum_{x,y=1}^{m}M_{xy}\langle a_x,b_y\rangle,
  \label{eq:SDPd-matrix}
\end{equation}
where the maximum is taken over all unit vectors $a_1,\ldots,a_m$ and
$b_1,\ldots,b_m$ in $\mathbb{R}^{d}$. For $d=1$, unit vectors reduce to signs,
giving the binary value
\begin{equation}
  L(M):=Q_1(M)
  =
  \max_{a,b\in\{-1,+1\}^{m}}
  \sum_{x,y=1}^{m}M_{xy}a_xb_y.
  \label{eq:SDP1-matrix}
\end{equation}
Maximizing over $b$ while keeping $a$ fixed gives the equivalent expression
\begin{equation}
  L(M)
  =
  \max_{a\in\{-1,+1\}^{m}}
  \sum_{y=1}^{m}
  \left|\sum_{x=1}^{m}M_{xy}a_x\right|.
  \label{eq:SDP1-one-sided}
\end{equation}

The real Grothendieck constant of order $d$ is
\begin{equation}
  \KG(d)
  :=
  \sup_{M\neq0}\frac{Q_d(M)}{L(M)},
  \label{eq:finite-Grothendieck-constant}
\end{equation}
where the supremum is taken over nonzero real matrices of all finite sizes. It
is enough to consider square matrices, since rectangular matrices can be made
square by adding rows or columns of zeros. Equivalently, $\KG(d)$ is the
smallest constant for which $Q_d(M)\leq\KG(d)L(M)$ holds for every finite real
matrix. The sequence $\KG(d)$ is nondecreasing and bounded, and its limit is
the real Grothendieck constant:
\begin{equation}
  \KG=\lim_{d\to\infty}\KG(d).
  \label{eq:KG-limit}
\end{equation}
Here $d$ denotes the dimension of the real vectors; its relation to the local
Hilbert-space dimension of a quantum system is discussed in
Sec.~\ref{sec:physics}.

Apart from $\KG(1)=1$, the only known exact value is
$\KG(2)=\sqrt{2}$~\cite{Krivine1979}. The case $d=3$ has attracted particular
attention because of its connection to two-qubit Bell correlations. Successive
improvements have used explicit Bell inequalities and computational
optimization \cite{Vertesi08,Brierley2016,Hirsch2017,Tavakoli2020,
  Marton2025,Designolle2023,Designolle2026}.
Rigorous bounds place its value in the interval
\begin{equation*}
  1.43670\leq\KG(3)\leq1.4546.
\end{equation*}
A complementary machine-learning approach by von Selzam and Marquardt gives
the numerical estimate $p_c\simeq0.691$ for the critical visibility of
two-qubit Werner states under projective
measurements~\cite{vonSelzamMarquardt2025}. Through the relation
$p_c=1/\KG(3)$~\cite{Acin2006}, this suggests $\KG(3)\simeq1.447$, but does
not provide a rigorous bound.

For higher finite orders, notable constructions include those of Fishburn and
Reeds~\cite{FishburnReeds1994}, Divi\'anszky et al.~\cite{Divianszky2017}, and
Designolle et al.~\cite{Designolle2026}. The latter work gives certified
improvements using rectangular matrices for $d\in\{3,4,5\}$; since the
constants are nondecreasing, these also improve the previously known lower
bounds through $d=9$. The same work provides further candidates based on
heuristic optimization.

For the infinite-order constant, the Davie--Reeds lower bound
$\KG\geq1.676956674\ldots$~\cite{Davie1984,Reeds1991} remained unsurpassed for
more than four decades. Krivine's classical upper bound is
\begin{equation}
  \KG\leq
  \frac{\pi}{2\log(1+\sqrt{2})}
  \simeq1.782214.
\end{equation}
Braverman et al.~\cite{Braverman2013} proved that this upper bound is not
tight, and subsequent work gave larger explicit
improvements~\cite{HeilmanUpper2026,Saha2026}. Recent progress on the lower
bound includes the perturbative constructions of
Heilman~\cite{HeilmanLower2026} and Jones and
Malavolta~\cite{JonesMalavolta2026}. Saha et al.~\cite{Saha2026} obtained the
interval
\begin{equation}
  \frac{6\pi}{11}
  \simeq1.713596
  \leq\KG
  <1.781867.
  \label{eq:current-KG-interval}
\end{equation}
These advances motivate the search for explicit constructions that could yield
substantially larger lower bounds.

\subsection{Overview of our results}
\label{subsec:contributions}

We study a one-parameter family of rotationally invariant kernels,
\begin{equation}
  k_c(t)=t-ct^3,
  \qquad -1\leq t\leq1,
  \qquad c\geq0.
  \label{eq:cubic-kernel-intro}
\end{equation}
Here $t$ is the inner product of two unit vectors. We consider both vectors
distributed uniformly over the sphere and finite collections of generating
vectors. The cubic kernel provides an alternative to the Davie--Reeds
functional, which we review for comparison.

In both settings, the generating vectors give an explicitly attainable value
$\widehat Q_d(M_c)\leq Q_d(M_c)$. Consequently, whenever $L(M_c)>0$,
\begin{equation}
  \KG(d)
  \geq\frac{Q_d(M_c)}{L(M_c)}
  \geq\frac{\widehat Q_d(M_c)}{L(M_c)}.
  \label{eq:master-lower-bound}
\end{equation}
The numerator on the right can be evaluated directly. The main difficulty is
to bound the maximum binary value in the denominator.

We first analyze the continuous problem. The hemispherical binary strategies
assign opposite signs to the two hemispheres. We consider a small cubic
deformation of their boundaries and find that the quadratic change in the
binary value vanishes when
\begin{equation}
  c_*(d)=\frac{(d+1)(d+3)}{3(d+5)},
  \qquad d\geq3.
  \label{eq:cstar-gen}
\end{equation}
Above this value, the deformation increases the binary value, so the
hemispherical strategies are no longer locally optimal. This calculation does
not prove their optimality below the threshold or determine the globally
optimal strategy above it.

Motivated by this analysis and our numerical results, we conjecture that the
hemispherical strategies are globally optimal throughout $0\leq c\leq c_*(d)$.
The corresponding candidate ratio increases with $c$ on this interval.
Evaluating it at the endpoint gives $\KG^{\mathrm{cub}}(d)$ in
Eq.~\eqref{eq:cubic-finite-bound} and, assuming the conjecture,
\begin{equation}
  \KG\geq
  \lim_{d\to\infty}\KG^{\mathrm{cub}}(d)
  =\frac{9\pi}{16}
  \simeq1.767146.
  \label{eq:main-asymptotic}
\end{equation}
To prove the proposed bound in dimension $d$, it suffices to show that no
binary strategy exceeds the value attained by the hemispherical strategies at
$c=c_*(d)$.

We then study finite discretizations in dimensions $d\in\{3,4,6,8,12,16,24\}$.
To estimate their binary optima, we use a large-scale see-saw
method~\cite{Araujo2020,Marton2024}, starting from different strategies and
applying additional perturbations. The largest binary value found satisfies
$L_{\mathrm{heur}}(M_c)\leq L(M_c)$. Replacing the denominator in
Eq.~\eqref{eq:master-lower-bound} by this heuristic value therefore produces
only a \emph{candidate lower bound}. Establishing the reported ratio requires a
matching upper bound on $L(M_c)$; a weaker upper bound may still yield a
useful rigorous lower bound or certify a dimension witness.

For $d=3,4,6,8$, Fibonacci-type and Sobol--Gaussian discretizations give
heuristic ratios that approach the continuous cubic predictions as the number
of vectors increases. Varying the kernel parameter also reveals departures
from hemispherical behavior near $c_*(d)$. Lattice-shell constructions extend
the numerical study to $d=24$. In particular, the Barnes--Wall shell of index
$n=6$ (squared norm $12$), with $m=4\,480\,320$ vectors after retaining one
representative from each opposite pair, gives
\begin{equation}
  \frac{\widehat Q_{16}(M_c)}
  {L_{\mathrm{heur}}(M_c)}
  \approx1.68204729.
\end{equation}
This candidate ratio exceeds the Davie--Reeds lower bound for $\KG$. The
numerical comparisons support the continuous predictions but do not certify
the binary optima.

Finally, we interpret the finite matrices as coefficients of bipartite
correlation Bell expressions. Combining our candidate ratios with an upper
bound on $\KG(7)$ yields candidate dimension witnesses with finitely many
measurement settings. If the required binary upper bounds are certified, the
corresponding violations establish Schmidt number at least five and hence
local Hilbert-space dimension at least five for each party.

\subsection{Organization of the paper}
\label{subsec:organization}

Section~\ref{sec:kernels} introduces the continuous formulation, reviews the
Davie--Reeds functional, and develops the cubic construction. It includes the
derivation of the critical parameter and the conjectured finite-order and
asymptotic bounds. Section~\ref{sec:discrete} presents the finite
discretizations, the numerical convergence tests, and the lattice-shell
results. Section~\ref{sec:physics} explains the connection to quantum
correlations and dimension witnesses. Section~\ref{sec:conclusion} discusses
the remaining certification problems and possible extensions.

Appendix~\ref{app:crit} gives the derivation of the critical parameter in
arbitrary dimension. Appendix~\ref{app:seesaw} describes the numerical
optimization and the efficient evaluation of large instances.
Appendix~\ref{app:const} provides the generating-vector constructions.

\section{Continuous formulations and kernel constructions}
\label{sec:kernels}

We now extend the matrix formulation to continuous sets of vectors. Let
$(X,\mu)$ and $(Y,\nu)$ be probability spaces, and let $M:X\times
Y\rightarrow\mathbb{R}$ be an integrable kernel. Following the notation
introduced in Eqs.~\eqref{eq:SDPd-matrix} and \eqref{eq:SDP1-matrix}, for
every integer $d\geq2$ we define
\begin{equation}
  \begin{aligned}
    Q_d(M)
    &:=
    \sup_{A,B}
    \int_X\!\int_Y
    M(x,y)\langle A(x),B(y)\rangle\\
    &\qquad\qquad\qquad
    \times\mathrm{d}\nu(y)\mathrm{d}\mu(x),
  \end{aligned}
  \label{eq:Qd-kernel-definition}
\end{equation}
where the supremum is taken over all measurable assignments of unit vectors,
\begin{equation}
  A:X\rightarrow S^{d-1},
  \qquad
  B:Y\rightarrow S^{d-1}.
\end{equation}
Similarly, we define
\begin{equation}
  \begin{aligned}
    L(M)
    &:=
    \sup_{a,b}
    \int_X\!\int_Y
    M(x,y)a(x)b(y)\\
    &\qquad\qquad\qquad
    \times\mathrm{d}\nu(y)\mathrm{d}\mu(x),
  \end{aligned}
  \label{eq:L-kernel-definition}
\end{equation}
where the supremum is taken over all measurable sign functions
\begin{equation}
  \begin{aligned}
    a&:X\rightarrow\{-1,+1\},\\
    b&:Y\rightarrow\{-1,+1\}.
  \end{aligned}
\end{equation}
As in the finite-matrix formulation of
Eq.~\eqref{eq:finite-Grothendieck-constant}, $\KG(d)$ is the supremum of
$Q_d(M)/L(M)$ over all probability spaces and all integrable real kernels $M$
satisfying $L(M)>0$. In this work, we consider the unit sphere $S^{d-1}$ and
set
\begin{equation}
  X=Y=S^{d-1},
  \qquad
  \mu=\nu=\sigma_d,
\end{equation}
where $\sigma_d$ is the uniform probability measure on $S^{d-1}$. Thus every
integrable real-valued kernel
\begin{equation}
  M:S^{d-1}\times S^{d-1}\rightarrow\mathbb{R}
\end{equation}
with $L(M)>0$ yields the lower bound
\begin{equation}
  \KG(d)
  \geq
  \frac{Q_d(M)}{L(M)}.
  \label{eq:KG-kernel-ratio}
\end{equation}
We first review the Davie--Reeds construction \cite{Davie1984,Reeds1991} and
then introduce our cubic kernel.

\subsection{The Davie--Reeds kernel}
\label{sec:Davie-Reeds}

For $0<\rho<1$, define the Davie--Reeds binary functional by
\begin{equation}
  \begin{aligned}
    &\mathcal{M}_{\rho}(a,b)\\
    &\quad:=
    d\int\!\!\int
    \langle x,y\rangle a(x)b(y)\,
    \mathrm{d}\sigma_d(x)\mathrm{d}\sigma_d(y)\\
    &\qquad
    -\rho\int
    a(x)b(x)\,\mathrm{d}\sigma_d(x).
  \end{aligned}
  \label{eq:Davie-functional}
\end{equation}
Here and below, integrals without specified domains are taken over $S^{d-1}$
with respect to normalized spherical measure. In distributional notation, the
corresponding kernel is
\begin{equation}
  M_{\rho}(x,y)
  =
  d\langle x,y\rangle-\rho\,\delta_x(y),
  \label{eq:Davie-kernel}
\end{equation}
where the Dirac term represents the single integral in
Eq.~\eqref{eq:Davie-functional}. Although this kernel is not an ordinary
integrable function, the construction can be justified by dividing the sphere
into small cells. Restricting the binary strategies to be constant within each
cell gives a finite matrix whose binary optimum cannot exceed that of the
continuous functional. Choosing one representative unit vector per cell gives
a vector value approaching $1-\rho$ as the cells shrink. Thus the
finite-matrix definition implies $\KG(d)\geq(1-\rho)/L(M_\rho)$.

For the natural vector assignment
\begin{equation}
  A(x)=x,
  \qquad
  B(y)=y,
\end{equation}
the resulting value is
\begin{equation}
  \begin{aligned}
    &\widehat Q_d(M_{\rho})\\
    &\quad=
    d\int\!\!\int
    \langle x,y\rangle^2\,
    \mathrm{d}\sigma_d(x)\mathrm{d}\sigma_d(y)
    -\rho\\
    &\quad=1-\rho,
  \end{aligned}
  \label{eq:Davie-natural-value}
\end{equation}
where we used the identity~\cite{Folland2001}
\begin{equation}
  \int\!\!\int
  \langle x,y\rangle^2\,
  \mathrm{d}\sigma_d(x)\mathrm{d}\sigma_d(y)
  =
  \frac{1}{d}.
\end{equation}
Consequently,
\begin{equation}
  Q_d(M_{\rho})
  \geq
  \widehat Q_d(M_{\rho})
  =
  1-\rho.
  \label{eq:Davie-quantum-value}
\end{equation}

Recall that the mean absolute value of a coordinate on the uniform sphere is
\begin{equation}
  \begin{aligned}
    \mu_d
    &:=
    \int_{S^{d-1}}|x_1|\,\mathrm{d}\sigma_d(x)\\
    &=
    \frac{\Gamma(d/2)}
    {\sqrt{\pi}\,\Gamma((d+1)/2)}.
  \end{aligned}
  \label{eq:mu-d-Davie}
\end{equation}
For $0\leq\lambda\leq1$, define
\begin{equation}
  g_d(\lambda)
  :=
  \mu_d(1-\lambda^2)^{(d-1)/2}
  \label{eq:gd-lambda}
\end{equation}
and
\begin{equation}
  \begin{aligned}
    h_d(\lambda)
    &:=1-2(d-1)\mu_d\\
    &\qquad\times
    \int_{\lambda}^{1}
    (1-t^2)^{(d-3)/2}\,\mathrm{d}t.
  \end{aligned}
  \label{eq:hd-lambda}
\end{equation}

The Davie--Reeds threshold argument reduces the binary optimization to
\begin{equation}
  L(M_{\rho})
  =
  \max_{0\leq\lambda\leq1}
  \left[
  d\,g_d(\lambda)^2
  +\rho\,h_d(\lambda)
  \right].
  \label{eq:Davie-classical-value}
\end{equation}
If the maximum occurs inside the interval, the derivative with respect to
$\lambda$ vanishes, giving
\begin{equation}
  \rho
  =
  \rho_{\mathrm{DR}}(d,\lambda)
  :=
  d\lambda g_d(\lambda),
  \label{eq:rho-lambda-short}
\end{equation}
or, explicitly,
\begin{equation}
  \rho_{\mathrm{DR}}(d,\lambda)
  =
  d\mu_d\lambda
  (1-\lambda^2)^{(d-1)/2}.
  \label{eq:rho-lambda}
\end{equation}
The branch relevant to the lower-bound construction lies in
\begin{equation}
  0<\lambda<\frac{1}{\sqrt d}.
\end{equation}
Combining Eqs.~\eqref{eq:Davie-quantum-value} and
\eqref{eq:Davie-classical-value}, we obtain
\begin{equation}
  \begin{aligned}
    \KG(d)
    &\geq
    \sup_{0<\lambda<1/\sqrt d}\\
    &\quad
    \frac{
      1-\rho_{\mathrm{DR}}(d,\lambda)
    }{
      d\,g_d(\lambda)^2
      +
      \rho_{\mathrm{DR}}(d,\lambda)h_d(\lambda)
    }.
  \end{aligned}
  \label{eq:Davie-Reeds-finite-bound}
\end{equation}
For the dimensions considered here, Table~\ref{tab:comparison-bounds} lists
the values of $\rho_{\mathrm{DR}}(d,\lambda)$ that optimize this ratio,
together with the corresponding Davie--Reeds lower bounds.

\subsection{The cubic kernel}
\label{subsec:cubic-kernel}

We consider the rotationally invariant kernel
\begin{equation}
  \begin{gathered}
    M_c(x,y)=k_c(\langle x,y\rangle),\\
    k_c(t)=t-ct^3,
    \qquad c\geq0.
  \end{gathered}
  \label{eq:cubic-kernel}
\end{equation}
The cubic term is the simplest nonlinear odd polynomial correction to the
scalar-product kernel. In particular,
\begin{equation}
  M_c(-x,y)=M_c(x,-y)=-M_c(x,y).
  \label{eq:antipodal-oddness}
\end{equation}
In the discrete constructions of Sec.~\ref{sec:discrete}, this symmetry allows
us to retain only one representative from each pair of opposite generating
vectors.

For the natural vector assignment $A(x)=x$ and $B(y)=y$, the resulting value
is
\begin{equation}
  \begin{aligned}
    &\widehat Q_d(M_c)\\
    &\quad=
    \int\!\!\int
    \left[
    \langle x,y\rangle^2
    -c\langle x,y\rangle^4
    \right]
    \,\mathrm{d}\sigma_d(x)\mathrm{d}\sigma_d(y)\\
    &\quad=
    \frac{1}{d}
    \left(1-\frac{3c}{d+2}\right),
  \end{aligned}
  \label{eq:cubic-natural-value}
\end{equation}
where we used the second and fourth moments of normalized spherical
measure~\cite{Folland2001}. This explicit assignment gives
$Q_d(M_c)\geq\widehat Q_d(M_c)$.

For binary strategies, we define
\begin{equation}
  \begin{aligned}
    &\mathcal{L}_c(a,b)\\
    &\quad:=
    \int\!\!\int
    k_c(\langle x,y\rangle)a(x)b(y)
    \,\mathrm{d}\sigma_d(x)\mathrm{d}\sigma_d(y),
  \end{aligned}
  \label{eq:cubic-binary-functional}
\end{equation}
so that
\begin{equation}
  L(M_c)=\sup_{a,b}\mathcal{L}_c(a,b),
  \label{eq:cubic-binary-value}
\end{equation}
where the supremum is taken over measurable sign functions
$a,b:S^{d-1}\rightarrow\{-1,+1\}$.

A particularly simple choice consists of aligned hemispherical strategies,
which assign opposite signs to the two hemispheres using a common axis. By
rotational invariance, we may choose this axis along the last coordinate:
\begin{equation}
  \begin{aligned}
    a_{\mathrm{HS}}(x)&=\operatorname{sgn}(x_d),\\
    b_{\mathrm{HS}}(y)&=\operatorname{sgn}(y_d).
  \end{aligned}
  \label{eq:hemispherical-strategy}
\end{equation}
The choice of $\operatorname{sgn}(0)$ does not affect the integrals. Using the
spherical absolute moment $\mu_d$ defined in Eq.~\eqref{eq:mu-d-Davie}, we
obtain
\begin{equation}
  \begin{aligned}
    &\int\!\!\int
    \langle x,y\rangle
    a_{\mathrm{HS}}(x)b_{\mathrm{HS}}(y)
    \,\mathrm{d}\sigma_d(x)\mathrm{d}\sigma_d(y)\\
    &\qquad=\mu_d^2,
  \end{aligned}
  \label{eq:hemisphere-linear}
\end{equation}
\begin{equation}
  \begin{aligned}
    &\int\!\!\int
    \langle x,y\rangle^3
    a_{\mathrm{HS}}(x)b_{\mathrm{HS}}(y)
    \,\mathrm{d}\sigma_d(x)\mathrm{d}\sigma_d(y)\\
    &\qquad=\mu_d^2\frac{3d+1}{(d+1)^2}.
  \end{aligned}
  \label{eq:hemisphere-cubic}
\end{equation}
These identities follow from standard spherical moment
formulas~\cite{Folland2001}. The hemispherical value is therefore
\begin{equation}
  L_{\mathrm{HS}}(d,c)
  =
  \mu_d^2
  \left[
  1-\frac{(3d+1)c}{(d+1)^2}
  \right].
  \label{eq:hemisphere-value}
\end{equation}

Our numerical results suggest that these strategies are globally optimal up to
a critical parameter $c_*(d)$. We next identify a family of small deformations
of the hemispherical boundaries that increases the binary value whenever
$c>c_*(d)$.

\subsection{Identifying the critical parameter}
\label{subsec:critical-parameter}

The calculation is most transparent in three dimensions, where the
perturbation is a threefold distortion of the equator. We give the complete
argument for this case and then state its extension to arbitrary $d\geq3$. The
general derivation is presented in Appendix~\ref{app:crit}.

\subsubsection{The case \texorpdfstring{$d=3$}{d=3}}
\label{sec:cstar-d3}

Write a point on $S^2$ as
\begin{equation}
  \begin{gathered}
    x=
    \left(
    \sqrt{1-u^2}\cos\phi,\,
    \sqrt{1-u^2}\sin\phi,\,
    u
    \right),\\
    \mathrm{d}\sigma_3(x)
    =
    \frac12\,\mathrm{d}u\,\frac{\mathrm{d}\phi}{2\pi}.
  \end{gathered}
  \label{eq:sphere-measure-d3}
\end{equation}
Here $u=x_3$, and angular brackets with a subscript $\phi$ will denote
averaging over $\phi\in[0,2\pi[$. The unperturbed strategies are
$a_0(x)=b_0(x)=\operatorname{sgn}(u)$. Since $\mu_3=1/2$, their value is
\begin{equation}
  L_{\mathrm{HS}}(3,c)
  =
  \frac14-\frac{5c}{32}.
  \label{eq:LHS-d3}
\end{equation}

Consider the cubic polynomial
\begin{equation}
  H(x)=x_1^3-3x_1x_2^2
  =(1-u^2)^{3/2}\cos(3\phi),
  \label{eq:H-d3}
\end{equation}
and deform the two strategies in opposite directions:
\begin{align}
  a_s(x)&=\operatorname{sgn}\!\left[u+sH(x)\right],
  \notag\\
  b_s(x)&=\operatorname{sgn}\!\left[u-sH(x)\right].
  \label{eq:perturbed-strategies-d3}
\end{align}
For sufficiently small $|s|$, their sign boundaries are
\begin{align}
  u_a(\phi)&=-s h(\phi)+O(s^3),
  \notag\\
  u_b(\phi)&=+s h(\phi)+O(s^3),
  \qquad h(\phi)=\cos(3\phi).
  \label{eq:boundaries-d3}
\end{align}
Thus $s$ controls the amplitude of a threefold deformation of the equator.

To calculate the change in the objective, first consider the field generated
by the undeformed hemisphere:
\begin{align}
  F_c(u)
  &:=
  \int_{S^2}
  k_c(\langle x,y\rangle)
  \operatorname{sgn}(y_3)\,
  \mathrm{d}\sigma_3(y)
  \notag\\
  &=
  \frac{u}{8}\left(4-3c+cu^2\right).
  \label{eq:hemisphere-field-d3}
\end{align}
The second equality follows by expanding the cubic term and using
\begin{equation}
  \begin{gathered}
    \int |y_3|\,\mathrm{d}\sigma_3=\frac12,
    \qquad
    \int |y_3|^3\,\mathrm{d}\sigma_3=\frac14,\\
    \int |y_3|y_1^2\,\mathrm{d}\sigma_3
    =
    \int |y_3|y_2^2\,\mathrm{d}\sigma_3
    =\frac18.
  \end{gathered}
\end{equation}

Let $\delta a=a_s-a_0$ and $\delta b=b_s-b_0$. By bilinearity and symmetry of
the kernel,
\begin{align}
  &\mathcal{L}_c(a_s,b_s)-L_{\mathrm{HS}}(3,c)
  \notag\\
  &\quad=
  \int F_c(u)\bigl[\delta a(x)+\delta b(x)\bigr]
  \,\mathrm{d}\sigma_3(x)
  \notag\\
  &\qquad+
  \int\!\!\int
  k_c(\langle x,y\rangle)\delta a(x)\delta b(y)
  \,\mathrm{d}\sigma_3(x)\mathrm{d}\sigma_3(y).
  \label{eq:variation-splitting-d3}
\end{align}
The first term describes the effect of moving each boundary while the other
party remains hemispherical. The second accounts for the interaction between
the two deformations.

For Alice, shifting the boundary from $0$ to $u_a(\phi)$ gives
\begin{align}
  \int F_c(u)\delta a(x)\,\mathrm{d}\sigma_3(x)
  &=
  -\left\langle
  \int_0^{u_a(\phi)}F_c(u)\,\mathrm{d}u
  \right\rangle_\phi
  \notag\\
  &=
  -\frac{4-3c}{32}s^2+O(s^4),
  \label{eq:single-boundary-d3}
\end{align}
where we used $\langle h^2\rangle_\phi=1/2$. Bob's contribution is identical.

At leading order, the mixed term is obtained by evaluating the kernel on the
equator:
\begin{align}
  &\int\!\!\int
  k_c(\langle x,y\rangle)\delta a(x)\delta b(y)
  \,\mathrm{d}\sigma_3(x)\mathrm{d}\sigma_3(y)
  \notag\\
  &\quad=
  -s^2
  \left\langle
  h(\phi)h(\psi)
  k_c\!\left(\cos(\phi-\psi)\right)
  \right\rangle_{\phi,\psi}
  +O(s^4).
  \label{eq:mixed-boundary-d3}
\end{align}
The elementary identity
\begin{equation}
  k_c(\cos\theta)
  =
  \left(1-\frac{3c}{4}\right)\cos\theta
  -\frac{c}{4}\cos(3\theta)
\end{equation}
and orthogonality of the angular modes give
\begin{equation}
  \left\langle
  h(\phi)h(\psi)
  k_c\!\left(\cos(\phi-\psi)\right)
  \right\rangle_{\phi,\psi}
  =
  -\frac{c}{16}.
\end{equation}
The mixed contribution is therefore $cs^2/16+O(s^4)$.

Combining the two boundary contributions with the mixed term yields
\begin{equation}
  \mathcal{L}_c(a_s,b_s)
  =
  L_{\mathrm{HS}}(3,c)
  +\frac{c-1}{4}s^2
  +O(s^4).
  \label{eq:second-variation-d3}
\end{equation}
Odd powers of $s$ vanish because changing $s$ to $-s$ exchanges Alice and Bob
and leaves the objective unchanged.

For $c<1$, sufficiently small nonzero perturbations of this form decrease the
value. At $c=1$, the quadratic correction vanishes, whereas for $c>1$ it is
positive. Thus the threefold mode is marginal at quadratic order when
\begin{equation}
  c_*(3)=1,
  \qquad
  L_{\mathrm{HS}}(3,1)=\frac{3}{32}.
\end{equation}
In particular, the hemispherical pair cannot remain locally optimal for $c>1$.

\subsubsection{The result for arbitrary \texorpdfstring{$d\geq3$}{d ≥ 3}}
\label{sec:cstar-general-d}

The same deformation can be used in every dimension $d\geq3$. Taking $u=x_d$,
we retain the explicit polynomial $H(x)=x_1^3-3x_1x_2^2$ and set
\begin{align}
  a_s(x)&=\operatorname{sgn}\!\left[x_d+sH(x)\right],
  \notag\\
  b_s(x)&=\operatorname{sgn}\!\left[x_d-sH(x)\right].
  \label{eq:perturbed-strategies-general}
\end{align}
This polynomial is a harmonic cubic in the coordinates transverse to the
hemispherical axis.

The calculation in Appendix~\ref{app:crit} gives the critical parameter
\begin{equation}
  c_*(d)
  =
  \frac{(d+1)(d+3)}{3(d+5)}
  \label{eq:cstar-general}
\end{equation}
and the expansion
\begin{equation}
  \begin{split}
    \mathcal{L}_c(a_s,b_s)
    &=
    L_{\mathrm{HS}}(d,c)\\
    &\quad+
    \frac{24\mu_d^2}{(d+1)(d+3)}
    \left(\frac{c}{c_*(d)}-1\right)s^2
    +O(s^4).
  \end{split}
  \label{eq:second-variation-general}
\end{equation}
For $d=3$, this reduces to Eq.~\eqref{eq:second-variation-d3}.

Since the prefactor is positive, the deformation decreases the value for
$c<c_*(d)$ and increases it for $c>c_*(d)$. At $c=c_*(d)$, it is marginal at
quadratic order. The appendix also shows that the same threshold applies to
every transverse harmonic cubic.

This calculation proves that the hemispherical strategies are not locally
optimal above $c_*(d)$. It does not establish their optimality below this
value, nor does the vanishing quadratic term determine the effect of
higher-order terms at the threshold.

\subsection{Conjectured lower bounds}
\label{subsec:conjectured-cubic-bounds}

The preceding calculation, together with the numerical results in
Sec.~\ref{sec:discrete}, motivates the following conjecture.

\begin{conjecture}
  \label{conj:hemispherical-optimality}
  For every integer $d\geq3$, the hemispherical strategies in
  Eq.~\eqref{eq:hemispherical-strategy} are globally optimal for $L(M_c)$
  throughout the interval $0\leq c\leq c_*(d)$, where $c_*(d)$ is given in
  Eq.~\eqref{eq:cstar-general}. Equivalently,
  \begin{equation}
    L(M_c)=L_{\mathrm{HS}}(d,c),
    \qquad 0\leq c\leq c_*(d).
  \end{equation}
\end{conjecture}

Assuming this conjecture, the explicit vector assignment in
Eq.~\eqref{eq:cubic-natural-value} yields
\begin{equation}
  \KG(d)
  \geq
  \frac{\widehat Q_d(M_c)}{L_{\mathrm{HS}}(d,c)}
  =:\KG^{\mathrm{cub}}(d,c),
  \qquad 0\leq c\leq c_*(d).
  \label{eq:new-bound-general-c}
\end{equation}
Explicitly,
\begin{equation}
  \KG^{\mathrm{cub}}(d,c)
  =
  \frac{
    \pi\Gamma\!\left(\frac{d+1}{2}\right)^2
  }{
    d\Gamma\!\left(\frac d2\right)^2
  }
  \frac{
    1-\dfrac{3c}{d+2}
  }{
    1-\dfrac{(3d+1)c}{(d+1)^2}
  }.
  \label{eq:new-bound-gamma}
\end{equation}

For fixed $d\geq3$, this ratio increases throughout $0\leq c\leq c_*(d)$.
Indeed, its denominator is positive on this interval, and the sign of its
derivative is determined by
\begin{equation}
  \frac{3d+1}{(d+1)^2}-\frac{3}{d+2}
  =
  \frac{d-1}{(d+1)^2(d+2)}
  >0.
\end{equation}
The largest candidate value in this interval is therefore attained at
$c=c_*(d)$, giving the conditional bound
\begin{equation}
  \KG(d)\geq\KG^{\mathrm{cub}}(d),
  \label{eq:main-new-lower-bound}
\end{equation}
where
\begin{equation}
  \KG^{\mathrm{cub}}(d)
  :=
  \frac{
    \pi\Gamma\!\left(\frac{d+1}{2}\right)^2
  }{
    d\Gamma\!\left(\frac d2\right)^2
  }
  \frac{3(d+1)(3d+7)}{4(d+2)(2d+3)}.
  \label{eq:cubic-finite-bound}
\end{equation}
For this endpoint bound, it would suffice to prove the conjectured equality
$L(M_c)=L_{\mathrm{HS}}(d,c)$ at $c=c_*(d)$.

Taking the limit gives
\begin{equation}
  \lim_{d\to\infty}\KG^{\mathrm{cub}}(d)
  =
  \frac{9\pi}{16}
  \simeq1.767146.
  \label{eq:cubic-asymptotic-bound}
\end{equation}
Consequently, Conjecture~\ref{conj:hemispherical-optimality} would imply
\begin{equation}
  \KG\geq\frac{9\pi}{16}.
  \label{eq:infinite-new-bound}
\end{equation}

Table~\ref{tab:comparison-bounds} compares the Davie--Reeds lower bounds, the
cubic candidate values, and the Krivine upper bounds. In every dimension for
which a cubic candidate is listed, it exceeds the corresponding Davie--Reeds
bound. For $d=3$, however, the cubic candidate $64/45\simeq1.422222$ is below
the rigorous lower bound $1.43670$ of Ref.~\cite{Designolle2026}. In the
infinite-dimensional limit, the proposed value $9\pi/16$ would improve the
rigorous lower bound $6\pi/11\simeq1.713596$ reported in Ref.~\cite{Saha2026}.
These proposed improvements remain conditional on establishing the binary 
optimum rigorously.

\begin{table*}[t]
  \caption{Comparison of Davie--Reeds lower bounds, cubic candidate
    bounds, and Krivine upper bounds for the real Grothendieck constants. The
    row $d=\infty$ refers to $\KG$, and numerical entries are rounded to six
    decimal places. The Davie--Reeds parameter
    $\rho_{\mathrm{DR}}(d,\lambda_{\mathrm{opt}})$ maximizes the ratio in
    Eq.~\eqref{eq:Davie-Reeds-finite-bound}. The cubic candidates are
    evaluated at the critical parameter $c_*(d)$ of
    Eq.~\eqref{eq:cstar-general} and are conditional on hemispherical
    optimality at this parameter
    [Conjecture~\ref{conj:hemispherical-optimality}]. Boldface marks
    candidates that would improve previously reported lower bounds. For $d=3$,
    the cubic candidate $64/45\simeq1.422222$ is below the rigorous lower
    bound $1.43670$ of Ref.~\cite{Designolle2026}; a stronger upper bound,
    approximately $1.4546$, is given in Ref.~\cite{Designolle2023}. For
    comparison, Saha \emph{et al.}~\cite{Saha2026} report lower and upper
    bounds on $\KG$ of approximately $1.713596$ and $1.781867$, respectively.}
  \label{tab:comparison-bounds}
  \centering
  \small
  \renewcommand{\arraystretch}{1.15}
  \begin{ruledtabular}
    \begin{tabular}{cccccc}
      \shortstack{Dimension\\$d$}
      &
      \shortstack{Davie--Reeds parameter\\
        $\rho_{\mathrm{DR}}(d,\lambda_{\mathrm{opt}})$}
      &
      \shortstack{Davie--Reeds\\lower bound}
      &
      \shortstack{Critical parameter\\$c_*(d)$}
      &
      \shortstack{Cubic candidate\\$\KG^{\mathrm{cub}}(d)$}
      &
      \shortstack{Krivine\\upper bound}
      \\
      \hline
      2
      & 0.207740
      & 1.268417
      & --
      & --
      & 1.414214
      \\
      3
      & 0.205095
      & 1.386736
      & 1.000000
      & 1.422222
      & 1.516247
      \\
      4
      & 0.203522
      & 1.452471
      & 1.296296
      & \textbf{1.498315}
      & 1.570796
      \\
      5
      & 0.202484
      & 1.494108
      & 1.600000
      & \textbf{1.547253}
      & 1.606108
      \\
      6
      & 0.201748
      & 1.522787
      & 1.909091
      & \textbf{1.581281}
      & 1.631036
      \\
      7
      & 0.201199
      & 1.543722
      & 2.222222
      & \textbf{1.606275}
      & 1.649646
      \\
      8
      & 0.200775
      & 1.559667
      & 2.538462
      & \textbf{1.625391}
      & 1.664102
      \\
      9
      & 0.200438
      & 1.572212
      & 2.857143
      & \textbf{1.640475}
      & 1.675672
      \\
      10
      & 0.200163
      & 1.582338
      & 3.177778
      & \textbf{1.652675}
      & 1.685150
      \\
      11
      & 0.199934
      & 1.590682
      & 3.500000
      & \textbf{1.662744}
      & 1.693061
      \\
      12
      & 0.199742
      & 1.597675
      & 3.823529
      & \textbf{1.671192}
      & 1.699768
      \\
      16
      & 0.199201
      & 1.617095
      & 5.126984
      & \textbf{1.694686}
      & 1.718788
      \\
      24
      & 0.198644
      & 1.636785
      & 7.758621
      & \textbf{1.718540}
      & 1.738755
      \\
      $\infty$
      & 0.197479
      & 1.676957
      & $c_*(d)\sim d/3$
      & \textbf{1.767146}
      & 1.782214
      \\
    \end{tabular}
  \end{ruledtabular}
\end{table*}

\section{Discrete one-parameter functionals}
\label{sec:discrete}

In this section, we investigate finite discretizations of the cubic kernel and
compare their numerical behavior with the continuous predictions obtained
above. For a fixed dimension $d$, we choose $m$ unit vectors in $\mathbb{R}^d$
and construct the corresponding one-parameter family of $m\times m$ matrices
$M_c$. The unit vectors yield an explicit value $\widehat Q_d(M_c)$ for the
$d$-dimensional vector optimization, whereas evaluating the binary optimum
$L(M_c)$ requires optimization over signs $\pm1$. We approximate this optimum
using a large-scale see-saw heuristic and denote the largest value found by
$L_{\mathrm{heur}}(M_c)\leq L(M_c)$. Although the exact ratio $\widehat
Q_d(M_c)/L(M_c)$ is a lower bound on $K_G(d)$, replacing its denominator by
$L_{\mathrm{heur}}(M_c)$ gives only a candidate lower bound unless the binary
optimum is independently certified.

To compare the discrete and continuous problems, we fix $c=c_*(d)$ and
increase $m$ at fixed $d$. If the empirical distribution of the chosen vectors
converges to the normalized rotation-invariant surface measure on $S^{d-1}$,
then the exact discrete ratio converges to the corresponding continuous ratio.
Here the discrete objective values are normalized by $m^2$ when compared with
continuum integrals; this common factor cancels in their ratio. This
convergence does not require a particular lattice construction, but increasing
the number of vectors alone is insufficient. Our numerical study has two
parts. First, we consider increasingly fine spherical discretizations in
dimensions $d=3,4,6,8$ and compare the resulting heuristic ratios with the
continuous cubic predictions. Second, we examine structured point sets
obtained from lattice shells in dimensions $d=8,12,16,24$. The results provide
numerical support for the continuous predictions. Details of the see-saw
algorithm and the vector constructions are given in
Appendices~\ref{app:seesaw} and \ref{app:const}, respectively. All numerical
computations were performed on an HP Z8 G5 workstation equipped with 64
physical cores and 1 TB of DDR5 memory.

\subsection{Finite discretization using \texorpdfstring{$d$}{d}-dimensional unit vectors}
\label{subsec:discrete-construction}

Let $v_1,\ldots,v_m\in \R^d$ be a finite collection of unit vectors, and let
$G\in\mathbb{R}^{m\times m}$ denote their Gram matrix,
\begin{align}
  G_{xy}=\langle v_x,v_y\rangle.
  \label{eq:discrete-Gram}
\end{align}
For $c\geq0$, we associate with this collection the symmetric matrix with
entries
\begin{align}
  (M_c)_{xy}=G_{xy}-cG_{xy}^{3}.
  \label{eq:discrete-Mc}
\end{align}
The case $c=0$ recovers the linear kernel and will serve as a reference below.
In accordance with Eq.~\eqref{eq:SDP1-matrix}, the binary value of $M_c$ is
\begin{align}
  L(M_c)
  =
  \max_{a,b\in\{-1,+1\}^{m}}
  \sum_{x,y=1}^{m}(M_c)_{xy}a_xb_y.
  \label{eq:discrete-local-value}
\end{align}

The vectors themselves provide a feasible strategy for the $d$-dimensional
vector optimization: we set $A_x=v_x$ and $B_y=v_y$. The resulting value is
\begin{equation}
  \begin{aligned}
    &\widehat Q_d(M_c)
    :=
    \sum_{x,y=1}^{m}(M_c)_{xy}\langle v_x,v_y\rangle\\
    &\qquad=
    \sum_{x,y=1}^{m}G_{xy}^{2}
    -
    c\sum_{x,y=1}^{m}G_{xy}^{4}.
  \end{aligned}
  \label{eq:discrete-Qhat}
\end{equation}
Since this strategy need not maximize the vector objective, $\widehat
Q_d(M_c)\leq Q_d(M_c)$. Consequently, whenever $L(M_c)>0$,
\begin{align}
  K_G(d)
  \geq
  \frac{\widehat Q_d(M_c)}{L(M_c)}.
  \label{eq:discrete-KG-bound}
\end{align}

The computationally demanding quantity in Eq.~\eqref{eq:discrete-KG-bound} is
the binary optimum $L(M_c)$. For a fixed $b$, maximization over $a$ can be
performed explicitly, giving
\begin{align}
  L(M_c)
  =
  \max_{b\in\{-1,+1\}^{m}}
  \sum_{x=1}^{m}\bigl|(M_cb)_x\bigr|.
  \label{eq:discrete-local-reduced}
\end{align}
Direct enumeration therefore involves $2^m$ choices of $b$. Branch-and-bound
methods can substantially reduce this search, but their practical performance
depends strongly on the structure and numerical representation of the matrix.
For generic dense instances, even matrices of order roughly $100$ can be
challenging; here the matrix order is $m$, not the vector dimension $d$
\cite{Divianszky2017,Marton2025,Designolle2023,
  Designolle2026}.
The constructions considered below extend to several million vectors, so we
use the large-scale heuristic described in Appendix~\ref{app:seesaw}. Matrix
products and the evaluation of $\widehat Q_d(M_c)$ use the implicit formulas
in that appendix, without storing the full matrices $G$ or $M_c$.

The heuristic is based on alternating maximization of the objective in
Eq.~\eqref{eq:discrete-local-value}. Starting from a pair of binary vectors,
one first optimizes all components of $a$ with $b$ fixed, and then all
components of $b$ with $a$ fixed:
\begin{align}
  a_x^{(t+1)}
  &=
  \operatorname{sgn}\!\left(
  \sum_{y=1}^{m}(M_c)_{xy}b_y^{(t)}
  \right),
  \notag\\
  b_y^{(t+1)}
  &=
  \operatorname{sgn}\!\left(
  \sum_{x=1}^{m}(M_c)_{xy}a_x^{(t+1)}
  \right).
  \label{eq:discrete-seesaw-updates}
\end{align}
We take $\operatorname{sgn}(0)=1$, as in Appendix~\ref{app:seesaw}. Each
noiseless update exactly maximizes the objective over one block of variables
and therefore cannot decrease its value in exact arithmetic. A fixed point is
a pair for which neither block can improve the objective while the other is
held fixed, but it need not be globally optimal. In the implementation, the
updates stop when the signs cease to change, the improvement falls below a
prescribed tolerance, or the iteration limit is reached. The latter two
stopping conditions do not guarantee an exact fixed point.

Repeating the search from different initial strategies and applying the
perturbations described in Appendix~\ref{app:seesaw} helps explore different
solutions. Unlike the noiseless updates, these perturbations may temporarily
decrease the objective. The largest objective value among the returned
strategies is retained as $L_{\mathrm{heur}}(M_c)$.

Every strategy returned by the heuristic is feasible, so
\begin{align}
  L_{\mathrm{heur}}(M_c)\leq L(M_c).
  \label{eq:discrete-heuristic-value}
\end{align}
Thus, replacing $L(M_c)$ by $L_{\mathrm{heur}}(M_c)$ in
Eq.~\eqref{eq:discrete-KG-bound} does not, by itself, yield a rigorous lower
bound on $K_G(d)$. We refer to the resulting ratios $\widehat
Q_d(M_c)/L_{\mathrm{heur}}(M_c)$ as candidate or heuristic lower bounds.
Below, we assess their numerical stability and agreement with the continuous
predictions. These comparisons provide evidence for the performance of the
heuristic, but do not establish global optimality.

\subsection{Numerical validation in dimensions up to eight}
\label{subsec:discrete-validation}

We consider dimensions $d=3,4,6,8$, using point sets containing up to
approximately one million vectors. We compare the resulting heuristic ratios
with the continuous predictions from Sec.~\ref{sec:kernels}. For the
convergence study, we fix $c=c_*(d)$ in Eq.~\eqref{eq:discrete-Mc} and
increase $m$ at fixed $d$. The vector constructions are described in
Appendix~\ref{app:const}, and the MATLAB scripts generating them are provided
as auxiliary files with the arXiv submission.

For $d=3$, we use a spherical Fibonacci spiral \cite{Gonzalez2010}, with
vectors $v_1,\ldots,v_m\in \R^3$ defined in Appendix~\ref{app:const_d34}. For
$d=4$, we use the super-Fibonacci construction on $S^3$~\cite{Alexa2022},
described in Appendix~\ref{app:const_d34}. In both dimensions, the sizes
considered range from $m=20$ to $m=10^6$. The corresponding cubic parameters
are $c_*(3)=1$ and $c_*(4)=35/27$.

For $d=6$ and $d=8$, we use the Sobol--Gaussian constructions described in
Appendix~\ref{app:const_d68}, with $c_*(6)=21/11$ and $c_*(8)=33/13$,
respectively. Motivated by the base-two structure of Sobol sequences, we use
sizes $m=2^n$, with selected values of $n$ between $10$ and $20$. Here $m$
counts the retained vectors after the first two Sobol points are omitted, as
explained in Appendix~\ref{app:const_d68}. The transformation in
Eq.~\eqref{eq:sobol-sphere-vectors} maps uniformly distributed cube points to
vectors distributed according to the normalized rotation-invariant surface
measure on $S^{d-1}$. Thus, the asymptotic equidistribution of the Sobol
points yields the corresponding equidistribution on the sphere.

These constructions provide spherical quadrature sequences for comparing the
discrete and continuous problems. As discussed above, asymptotic
equidistribution ensures convergence of the exact discrete ratios to the
corresponding continuous ratios. The numerical comparisons below use heuristic
binary values and therefore provide a consistency check rather than a proof of
the predicted continuous optima.

\begin{table}[t]
  \centering
  \small
  \setlength{\tabcolsep}{4pt}
  \renewcommand{\arraystretch}{1.1}
  \begin{tabular}{clccc}
    \hline\hline
    \noalign{\vskip 2pt}
    $d$ & $m$
    & $\dfrac{L_0(M_c)}{L_{\mathrm{heur}}(M_c)}$
    & \shortstack{Cubic candidate\\$K_G^{\mathrm{cub}}(d)$}
    & $\dfrac{\widehat Q_d(M_c)}{L_{\mathrm{heur}}(M_c)}$
    \\[4pt]
    \hline
    \multirow{6}{*}{$3$}
    & $20$   & $0.921631457571$
    & \multirow{6}{*}{$1.422222222222$}
    & $1.295149253766$ \\
    & $10^2$ & $0.999355937908$ & & $1.407434625685$ \\
    & $10^3$ & $0.999943008518$ & & $1.421421024849$ \\
    & $10^4$ & $0.999999125576$ & & $1.422174862427$ \\
    & $10^5$ & $0.999999997877$ & & $1.422219025124$ \\
    & $10^6$ & $0.999999998915$ & & $1.422222044099$ \\
    \hline
    \multirow{7}{*}{$4$}
    & $20$   & $0.709991486502$
    & \multirow{7}{*}{$1.498315298815$}
    & $0.922702622857$ \\
    & $50$   & $0.977805473148$ & & $1.328583248741$ \\
    & $10^2$ & $0.983619952898$ & & $1.426646634812$ \\
    & $10^3$ & $0.999008689592$ & & $1.492644361129$ \\
    & $10^4$ & $0.999940214420$ & & $1.497699091048$ \\
    & $10^5$ & $0.999997301589$ & & $1.498242543111$ \\
    & $10^6$ & $0.999999845622$ & & $1.498309156810$ \\
    \hline
    \multirow{5}{*}{$6$}
    & $2^{10}$ & $0.966096654890$
    & \multirow{5}{*}{$1.581281222708$}
    & $1.485418076247$ \\
    & $2^{13}$ & $0.997084043103$ & & $1.565295152230$ \\
    & $2^{15}$ & $0.998945863266$ & & $1.574693069844$ \\
    & $2^{18}$ & $0.999935505211$ & & $1.579736622579$ \\
    & $2^{20}$ & $0.999989458730$ & & $1.580668641937$ \\
    \hline
    \multirow{5}{*}{$8$}
    & $2^{10}$ & $0.908145996739$
    & \multirow{5}{*}{$1.625390646289$}
    & $1.406630825224$ \\
    & $2^{13}$ & $0.987191531016$ & & $1.581500422099$ \\
    & $2^{15}$ & $0.998206774105$ & & $1.612845370416$ \\
    & $2^{18}$ & $0.999750293170$ & & $1.617360987847$ \\
    & $2^{20}$ & $0.999828168109$ & & $1.623425605929$ \\
    \hline\hline
  \end{tabular}
  \caption{
    Finite discretizations at $c=c_*(d)$, with $c_*(3)=1$, $c_*(4)=35/27$,
    $c_*(6)=21/11$, and $c_*(8)=33/13$. The point sets are spherical Fibonacci
    spirals for $d=3$, super-Fibonacci constructions for $d=4$, and
    Sobol--Gaussian constructions for $d=6,8$. The third column compares the
    linear-kernel reference value defined in Eq.~\eqref{eq:L0-definition} with
    the largest binary value found for $M_c$. The fourth column gives the
    continuous cubic prediction from Eq.~\eqref{eq:cubic-finite-bound}, shown
    once for each dimension. The last column gives the candidate lower bound
    obtained using the heuristic binary value. Decimal entries are truncated
    to $12$ decimal places; the displayed precision does not establish the
    binary optima.
  }
  \label{tab:discrete-convergence}
\end{table}

To assess whether a strategy found for the linear kernel remains competitive
after the cubic deformation, we introduce a reference value. Let
$a^{\mathrm{lin}},b^{\mathrm{lin}}\in\{-1,+1\}^{m}$ be a pair of binary
strategies obtained by applying the heuristic to $M_0=G$. We evaluate this
same pair on $M_c$:
\begin{align}
  L_0(M_c)
  &:=
  \sum_{x,y=1}^{m}(M_c)_{xy}
  a_x^{\mathrm{lin}}b_y^{\mathrm{lin}}
  \notag\\
  &=
  \sum_{x,y=1}^{m}(G_{xy}-cG_{xy}^{3})
  a_x^{\mathrm{lin}}b_y^{\mathrm{lin}}.
  \label{eq:L0-definition}
\end{align}
This reference strategy is feasible for every $c$, so $L_0(M_c)\leq L(M_c)$,
independently of whether it is optimal at $c=0$. We compare it with the
largest value found for the cubic matrix, $L_{\mathrm{heur}}(M_c)$. The ratio
$L_0(M_c)/L_{\mathrm{heur}}(M_c)$ measures the performance of the
linear-kernel reference relative to the best cubic-kernel strategy found.

Table~\ref{tab:discrete-convergence} shows two consistent trends. First,
$L_0(M_c)/L_{\mathrm{heur}}(M_c)$ approaches one in every dimension studied.
Thus, the linear-kernel reference and the best cubic-kernel strategy found
become increasingly close in objective value, although their sign assignments
need not coincide. Second, the candidate ratios $\widehat
Q_d(M_c)/L_{\mathrm{heur}}(M_c)$ approach the continuous predictions
$K_G^{\mathrm{cub}}(d)$ as $m$ increases. The agreement between the
finite-size ratios and the continuous predictions provides a numerical
consistency check on the overall discretization and optimization procedure,
but does not independently validate its individual components or establish the
binary optima rigorously.

The convergence is slower in the higher dimensions considered. For $d=6$,
increasing $m$ from $2^{10}$ to $2^{20}$ raises
$L_0(M_c)/L_{\mathrm{heur}}(M_c)$ from approximately $0.966097$ to $0.999989$,
while the candidate lower bound increases from $1.485418$ to $1.580669$, close
to the continuous prediction $K_G^{\mathrm{cub}}(6)\approx1.581281$. For
$d=8$, the corresponding reference ratio increases from approximately
$0.908146$ to $0.999828$, and the candidate lower bound rises from $1.406631$
to $1.623426$, compared with $K_G^{\mathrm{cub}}(8)\approx1.625391$. At the
largest size considered, the remaining discrepancy is therefore larger for
$d=8$ than for $d=6$.

\begin{figure}[t]
  \centering
  \includegraphics[width=1.05\columnwidth]{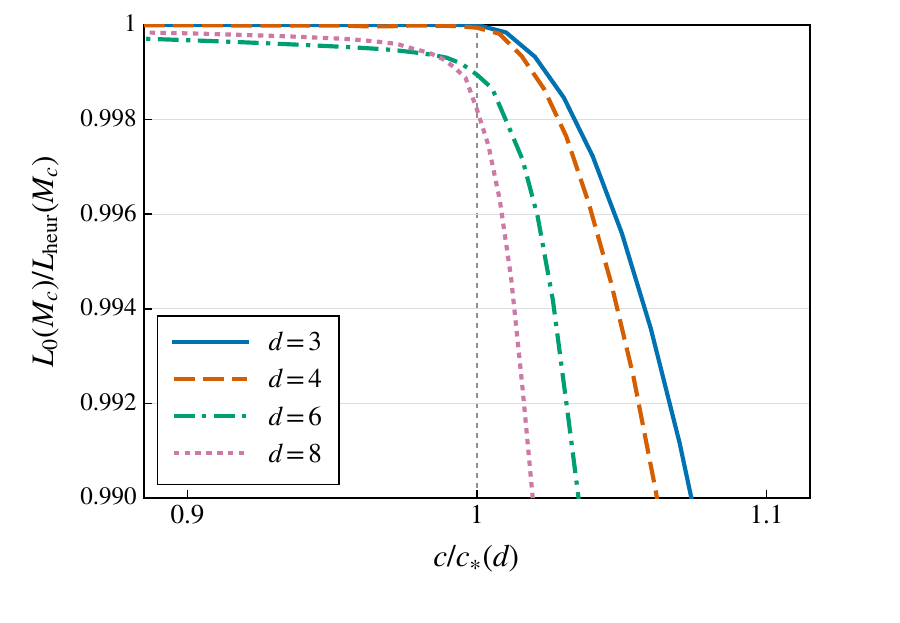}
  \caption{
    Performance of the linear-kernel reference near the predicted transition,
    measured by $L_0(M_c)/L_{\mathrm{heur}}(M_c)$ as a function of the
    normalized parameter $c/c_*(d)$. The curves correspond to $d=3,4,6,8$,
    with $c_*(3)=1$, $c_*(4)=35/27$, $c_*(6)=21/11$, and $c_*(8)=33/13$. We
    use $m=10^4$ vectors for $d=3,4$ and $m=2^{15}$ for $d=6,8$. The vertical
    dashed line marks the predicted threshold $c/c_*(d)=1$. The ratio remains
    close to one below the threshold and decreases beyond it, indicating that
    the heuristic finds strategies that outperform the reference.
  }
  \label{fig:cubic-breaking}
\end{figure}

We also examine how the reference strategy performs as $c$ varies around the
predicted threshold. Figure~\ref{fig:cubic-breaking} plots
$L_0(M_c)/L_{\mathrm{heur}}(M_c)$ against the normalized parameter $c/c_*(d)$
over the interval $[0.9,1.1]$. The $d=3,4$ curves use $m=10^4$ vectors,
whereas the $d=6,8$ curves use $m=2^{15}=32768$. In each dimension, the ratio
remains close to one below $c_*(d)$ and decreases beyond a transition region
near this value. The decrease shows that the search finds strategies that
outperform the fixed linear-kernel reference. This behavior is consistent with
the predicted departure from the hemispherical regime, although the ratio
alone does not determine the geometry of the strategies found. Discretization
effects and incomplete heuristic optimization can both affect the apparent
location of the transition.

\begin{figure}[t]
  \centering
  \includegraphics[width=0.9\linewidth]{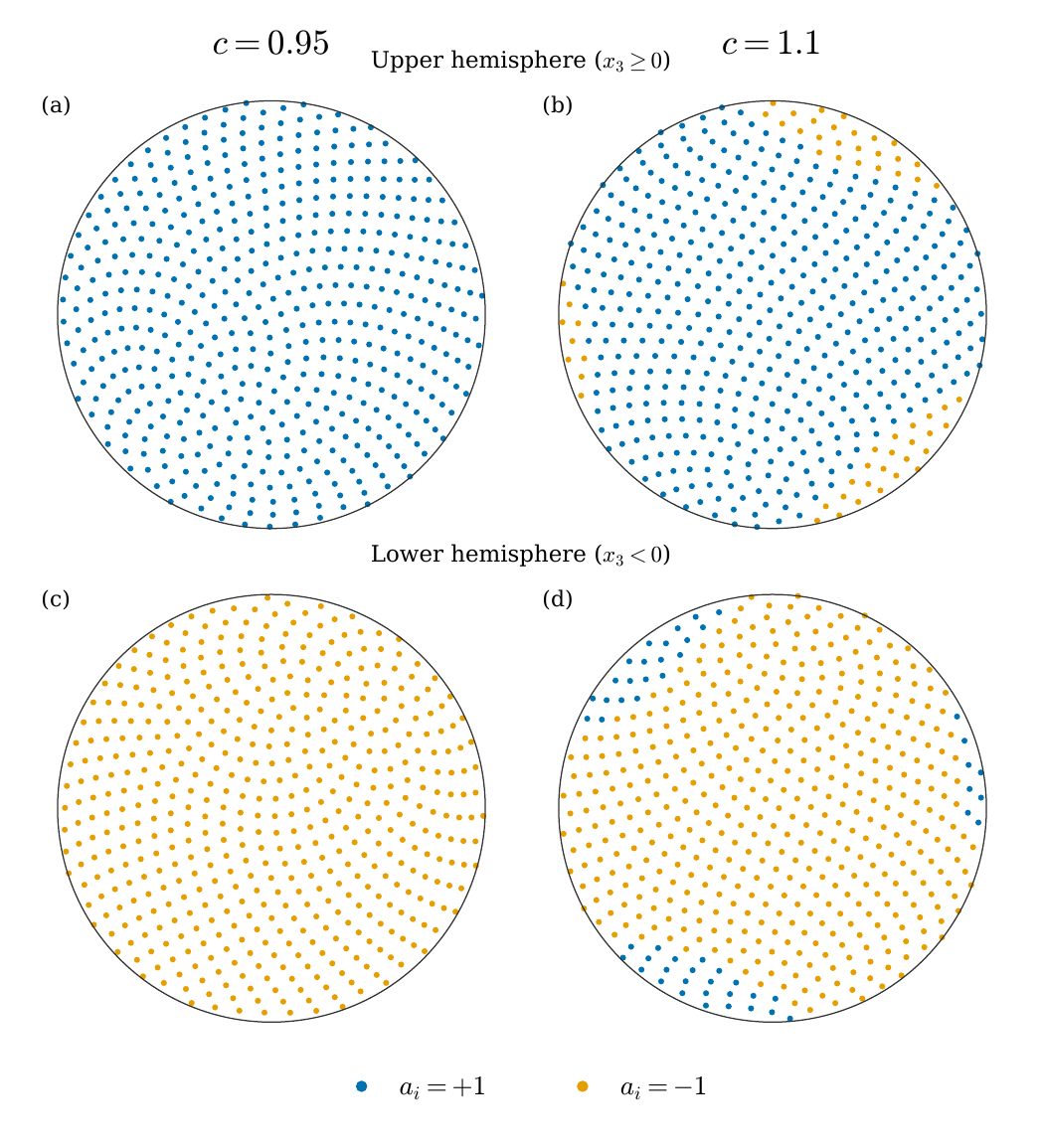}
  \caption{
    Binary strategies for Alice on a spherical Fibonacci grid with $d=3$ and
    $m=1000$ points. The left and right columns correspond to $c=0.95$ and
    $c=1.1$, respectively, below and above the predicted threshold $c_*(3)=1$.
    The upper ($x_3\geq0$) and lower ($x_3<0$) hemispheres appear in the top
    and bottom rows, using the equal-area projection
    $(X,Y)=(x_1,x_2)/\sqrt{1+|x_3|}$. Positive $x_1$ points right and positive
    $x_2$ points up in every panel; no reflection is applied to the lower
    hemisphere. Each circular boundary marks the equator. Blue and orange
    points represent $a_i=+1$ and $a_i=-1$, respectively. At $c=0.95$, the
    signs separate into two hemispheres. At $c=1.1$, the orange regions in the
    upper hemisphere and the blue regions in the lower hemisphere form a
    threefold pattern near the equator, consistent with a cubic-type
    deformation of the hemispherical strategy.
  }
  \label{fig:alice-strategies-d3}
\end{figure}

The geometry of the binary strategies is illustrated in
Fig.~\ref{fig:alice-strategies-d3} for a spherical Fibonacci grid with $d=3$
and $m=1000$ points. The upper and lower hemispheres are shown as equal-area
disks, with blue and orange points representing the signs $a_i=+1$ and
$a_i=-1$ returned by the heuristic. At $c=0.95<c_*(3)=1$ (left column), a
suitable rotation places all positive signs in the upper hemisphere and all
negative signs in the lower hemisphere. At $c=1.1>c_*(3)$ (right column),
orange regions in the upper hemisphere and blue regions in the lower
hemisphere form a threefold pattern near the equator. This pattern is
consistent with a cubic-type deformation of the hemispherical strategy beyond
the predicted transition.

\subsection{Lattice-shell constructions in dimensions up to \texorpdfstring{$24$}{24}}
\label{subsec:lattice-shells}

We now test the continuous predictions using symmetric point sets from four
lattices: $E_8$, the Coxeter--Todd lattice $K_{12}$, the Barnes--Wall lattice
$BW_{16}$, and the Leech lattice. Their dimensions are $8$, $12$, $16$, and
$24$, respectively.

A shell consists of all lattice vectors of a given length. Following the
convention of Appendix~\ref{app:const}, we define the shell of index $n$ by
\begin{align}
  \mathcal S_n(\Lambda)
  :=
  \{w\in\Lambda:\|w\|_2^2=2n\},
  \qquad n\geq1.
  \label{eq:lattice-shell-definition}
\end{align}
The shell index is therefore half the squared length; it need not equal the
ordinal number of a nonempty shell. Since $w$ and $-w$ always occur together,
we retain only the representative whose first nonzero coordinate is positive.
We then normalize these vectors:
\begin{equation}
  \begin{gathered}
    v_i=\frac{w_i}{\sqrt{2n}},
    \qquad i=1,\ldots,m,\\
    m=\frac{|\mathcal S_n(\Lambda)|}{2}.
  \end{gathered}
  \label{eq:lattice-shell-unit-vectors}
\end{equation}

Removing the antipodal copies leaves the exact ratio $\widehat
Q_d(M_c)/L(M_c)$ unchanged. Indeed, restoring $-v_i$ adds the negative of the
corresponding row and column of $M_c$, because the cubic kernel is odd. An
optimal binary strategy can therefore assign opposite signs to each antipodal
pair on both sides. Each original contribution is then repeated four times.
The same holds for the prescribed vector strategy, so restoring the full shell
multiplies both $L(M_c)$ and $\widehat Q_d(M_c)$ by four.

For each half-shell, we define $G$ and $M_c$ as in
Eqs.~\eqref{eq:discrete-Gram} and \eqref{eq:discrete-Mc}, with $c=c_*(d)$
given by Eq.~\eqref{eq:cstar-general}. We evaluate the candidate ratios
$\widehat Q_d(M_c)/L_{\mathrm{heur}}(M_c)$ using the same procedures as in the
previous subsection.

For $E_8$, we use shells with indices $n=1,5,15$, containing $m=120$,
$15\,120$, and $423\,360$ antipodal representatives, respectively. Their
construction and the shell-count formula are given in
Appendix~\ref{app:const_d8}. For $K_{12}$, we use shell indices $n=2,5,9$,
with squared lengths $4,10,18$. For $BW_{16}$, we use shell indices $n=2,4,6$,
with squared lengths $4,8,12$. For the Leech lattice, we use the first two
nonempty shells, with indices $n=2,3$ and squared lengths $4,6$. Its first
nonempty shell contains $196\,560$ vectors, giving $m=98\,280$ after antipodal
reduction. Details of the four constructions are given in
Appendices~\ref{app:const_d8}, \ref{app:const_d12}, \ref{app:const_d16}, and
\ref{app:const_d24}, respectively.

\begin{table*}[t]
  \centering
  {\small
    \setlength{\tabcolsep}{3.5pt}
    \begin{tabular}{lcccccc}
      \hline\hline
      Lattice
      & \shortstack{Shell index\\$n$}
      & $d$
      & $m$
      & $c_*(d)$
      & \shortstack{Discrete candidate\\
        $\widehat Q_d(M_c)/L_{\mathrm{heur}}(M_c)$}
      & \shortstack{Cubic candidate\\
        $K_G^{\mathrm{cub}}(d)$} \\[2pt]
      \hline
      $E_8$     & $1$  & $8$  & $120$
      & $33/13$  & $1.351416807943812$ & $1.625390646289$ \\
      $E_8$     & $5$  & $8$  & $15\,120$
      & $33/13$  & $1.613819278675966$ & $1.625390646289$ \\
      $E_8$     & $15$ & $8$  & $423\,360$
      & $33/13$  & $1.622121228586635$ & $1.625390646289$ \\
      \hline
      $K_{12}$  & $2$  & $12$ & $378$
      & $65/17$  & $0.940153472409670$ & $1.671191884731$ \\
      $K_{12}$  & $5$  & $12$ & $30\,240$
      & $65/17$  & $1.617111872073903$ & $1.671191884731$ \\
      $K_{12}$  & $9$  & $12$ & $510\,048$
      & $65/17$  & $1.658738133429911$ & $1.671191884731$ \\
      \hline
      $BW_{16}$ & $2$  & $16$ & $2\,160$
      & $323/63$ & $1.392803561679561$ & $1.694686204376$ \\
      $BW_{16}$ & $4$  & $16$ & $261\,360$
      & $323/63$ & $1.647233538966721$ & $1.694686204376$ \\
      $BW_{16}$ & $6$  & $16$ & $4\,480\,320$
      & $323/63$ & $1.682047299161634$ & $1.694686204376$ \\
      \hline
      Leech     & $2$  & $24$ & $98\,280$
      & $225/29$ & $1.637952291526756$ & $1.718539533613$ \\
      Leech     & $3$  & $24$ & $8\,386\,560$
      & $225/29$ & $1.680620166608739$ & $1.718539533613$ \\
      \hline\hline
    \end{tabular}
  }
  \caption{
    Candidate lower bounds from normalized lattice half-shells at $c=c_*(d)$.
    The index $n$ corresponds to squared length $2n$, as in
    Appendix~\ref{app:const}, and $m$ counts antipodal pairs. The penultimate
    column uses the heuristic binary value in the denominator, while the last
    column gives the continuous cubic prediction from
    Eq.~\eqref{eq:cubic-finite-bound}. The displayed precision does not
    certify the binary optima.
  }
  \label{tab:lattice-shell-bounds}
\end{table*}

For each lattice, the candidate ratios improve across the selected shells in
Table~\ref{tab:lattice-shell-bounds}. The $E_8$ shell of index $n=15$ gives
approximately $1.622121$, close to the continuous prediction $1.625391$. The
eighth nonempty $K_{12}$ shell, of index $n=9$, gives $1.658738$, compared
with $K_G^{\mathrm{cub}}(12)\approx1.671192$. The fifth nonempty $BW_{16}$
shell, of index $n=6$, gives $1.682047$, compared with
$K_G^{\mathrm{cub}}(16)\approx1.694686$.

For these lattices, the normalized full shells become uniformly distributed on
the sphere as the radius increases through the nonempty shells. This follows
from classical results on weighted theta series~\cite{Elkies2019}. Briefly,
these results show that averages of every nonconstant spherical harmonic over
a shell tend to zero. Since continuous functions on the sphere can be
approximated by finite sums of spherical harmonics, shell averages approach
uniform spherical averages.

It follows that the exact ratios $\widehat Q_d(M_c)/L(M_c)$ converge to the
corresponding continuous ratio at fixed $d$ and $c$. By the
antipodal-reduction argument above, this conclusion also applies to the ratios
computed from half-shells, although the retained representatives themselves
are not uniformly distributed over the full sphere. At $c=c_*(d)$, the limit
equals $K_G^{\mathrm{cub}}(d)$ if the proposed continuous binary optimum is
correct. This argument does not imply monotonic convergence from one shell to
the next, nor does it guarantee convergence of the heuristic ratios.

The second nonempty Leech shell, of index $n=3$, gives approximately
$1.680620$, slightly below the fifth nonempty $BW_{16}$ shell despite
containing more vectors. However, the number of points alone does not
determine how well a point set approximates the sphere. The Leech and
Barnes--Wall vectors lie on $S^{23}$ and $S^{15}$, respectively. The larger
gap to $K_G^{\mathrm{cub}}(24)\approx1.718540$ is consistent with finite-shell
effects. The present data do not separate these effects from errors due to
heuristic optimization.

Finally, the first $E_8$ shell shows that the continuous parameter $c_*(d)$
need not give the best candidate ratio for a finite point set. For its $120$
normalized antipodal representatives, the Gram matrix satisfies $G_{xx}=1$ and
$G_{xy}\in\{0,\pm1/2\}$ for $x\ne y$. Hence
\begin{align}
  \delta_{xy}=\frac{4G_{xy}^3-G_{xy}}{3},
\end{align}
where $\delta_{xy}$ is the Kronecker delta. Therefore,
\begin{equation}
  \begin{aligned}
    &G_{xy}-\frac{13}{6}\delta_{xy}\\
    &\qquad=
    \frac{31}{18}
    \left(G_{xy}-\frac{52}{31}G_{xy}^3\right).
  \end{aligned}
  \label{eq:E8-diagonal-cubic-equivalence}
\end{equation}
Thus, the cubic matrix at $c=52/31\approx1.677419$ is proportional to the
diagonal modification $G-(13/6)I$, where $I$ is the identity matrix. Table~I
of Ref.~\cite{Designolle2026} reports the candidate ratio
$165/109\approx1.513761$ for this matrix. Positive rescaling does not change
the ratio, so this candidate also applies to the cubic matrix at $c=52/31$. It
exceeds the value obtained here at $c=c_*(8)=33/13$ for the same first shell.
We therefore use $c_*(d)$ as a common reference for comparison with the
continuous problem, without claiming that it is optimal for each finite shell.

\section{Relation to quantum physics}
\label{sec:physics}

Measurements on entangled quantum systems can produce correlations that
violate Bell inequalities and therefore cannot be explained by local
hidden-variable models~\cite{Bell1964,Brunner2014}. Beyond detecting
nonlocality, suitably chosen inequalities can also place lower bounds on the
Hilbert-space dimension needed to reproduce the observed
statistics~\cite{Brunner2008}. These dimension witnesses require no detailed
characterization of the state or measurement devices. We first describe
dimension witnesses for bipartite quantum systems and their connection to
finite-order Grothendieck constants. We then present a finite-setting
candidate witness for local dimension greater than four.

\subsection{Dimension witnesses for bipartite quantum systems}

A dimension witness is a linear inequality satisfied by all quantum
realizations up to a given local dimension but violated by some
higher-dimensional realization~\cite{Brunner2008}. Here we consider two
parties, Alice and Bob, with measurement settings $x$ and $y$ and outcomes
$a,b\in\{-1,+1\}$. Their correlators are
\begin{equation}
  E_{xy}
  =
  \sum_{a,b=\pm1}ab\,p(a,b|x,y).
\end{equation}
For a real coefficient matrix $M=(M_{xy})$, every realization with both local
Hilbert-space dimensions at most $D$ satisfies
\begin{equation}
  \sum_{x,y}M_{xy}E_{xy}
  \leq W_D(M),
  \label{eq:general-dimension-witness}
\end{equation}
where
\begin{equation}
  \begin{split}
    W_D(M)
    &:=
    \max_{\rho,\{A_x\},\{B_y\}}\\
    &\quad
    \sum_{x,y}M_{xy}
    \operatorname{Tr}\!\left[\rho(A_x\otimes B_y)\right].
  \end{split}
  \label{eq:finite-dimensional-quantum-value}
\end{equation}
Here $\rho$ is a quantum state on $\mathbb{C}^{D}\otimes\mathbb{C}^{D}$: a
positive semidefinite Hermitian operator with $\operatorname{Tr}\rho=1$. The
operators $A_x$ and $B_y$ describe the binary measurements. They are Hermitian
and have eigenvalues in $[-1,1]$; for example, Alice's measurement effects are
$(\mathbb{I}_D\pm A_x)/2$. The resulting correlator is
$E_{xy}=\operatorname{Tr}[\rho(A_x\otimes B_y)]$.

To evaluate $W_D(M)$, it suffices to optimize over pure states
$\rho=\ket{\psi}\!\bra{\psi}$ and projective binary
measurements~\cite{Brunner2008}, for which
\begin{equation}
  A_x^2=B_y^2=\mathbb{I}_D.
\end{equation}
Indeed, for fixed measurements, the objective is linear in $\rho$, so its
maximum is attained at a pure state. Likewise, with the state and all other
measurements fixed, the objective is linear in each observable. Its maximum
over Hermitian operators with eigenvalues in $[-1,1]$ is attained at an
operator whose eigenvalues are all $\pm1$. Starting from an optimal
realization, we can therefore choose the state to be pure and each measurement
to be projective, without changing the optimum or increasing the local
dimension.

In fact, Eq.~\eqref{eq:general-dimension-witness} holds whenever the smaller
local dimension is at most $D$. By the Schmidt decomposition, any pure state
in such a realization has local supports of dimension at most $D$. Restricting
the observables to these supports preserves the correlators and gives
Hermitian operators with eigenvalues in $[-1,1]$, as allowed in the definition
of $W_D(M)$. Convexity extends the bound to mixed states and to realizations
using shared randomness. A violation therefore certifies local dimension at
least $D+1$ for each party.

For $D\geq2$, Lemma~2 of Ref.~\cite{Vertesi2009} connects this quantum bound
to the real-vector optimization defined in Eq.~\eqref{eq:SDPd-matrix}:
\begin{equation}
  W_D(M)\leq Q_{2D-1}(M).
  \label{eq:vertesi-pal-dimension-bound}
\end{equation}
In particular,
\begin{equation}
  W_4(M)\leq Q_7(M).
  \label{eq:vertesi-pal-dimension-bound47}
\end{equation}
Here $4$ is the local Hilbert-space dimension, whereas $7$ is the dimension of
the real vectors used to bound the quantum value.

To obtain a dimension witness, it suffices to find one explicit vector
strategy in some dimension $d'>7$ with value
\begin{equation}
  \widehat Q_{d'}(M)>Q_7(M).
\end{equation}
Tsirelson's construction realizes these vector correlations using binary
observables and a maximally entangled state of local dimension $D'=2^{\lfloor
d'/2\rfloor}$~\cite{Tsirelson1980,Tsirelson1987}. Denoting the value of this
particular quantum realization by $\widehat W_{D'}(M)$, we obtain
\begin{equation}
  \begin{aligned}
    W_4(M)
    &\leq Q_7(M)
    < \widehat Q_{d'}(M)\\
    &=\widehat W_{D'}(M).
  \end{aligned}
\end{equation}
Thus $M$ defines a dimension witness excluding realizations whose smaller
local dimension is at most four. One explicit realization suffices; neither
$Q_{d'}(M)$ nor $W_{D'}(M)$ needs to be determined.

\subsection{A finite-setting witness beyond local dimension four}

We combine the preceding dimension bound with Krivine's upper bound on
$K_G(7)$ to obtain a finite-setting candidate dimension witness. We use the
$m=4\,480\,320$ unit vectors $v_x\in\mathbb{R}^{16}$ from the Barnes--Wall
lattice $BW_{16}$ described in Appendix~\ref{app:const_d16}. We define $M=M_c$
as in Eq.~\eqref{eq:discrete-Mc}, with $c=c_*(16)=323/63$. The resulting
correlation inequality has $m$ binary measurement settings per party.

The vector strategy that assigns $v_x$ to Alice's setting $x$ and $v_y$ to
Bob's setting $y$ has value
\begin{align}
  \widehat Q_{16}(M)
  &=
  \sum_{x,y=1}^{m}M_{xy}\langle v_x,v_y\rangle
  \notag\\
  &=
  \frac{55m^2}{6048}
  =
  \frac{3\,833\,436\,464\,000}{21}.
  \label{eq:BW16-explicit-vector-value}
\end{align}
Here we used the shell's second- and fourth-moment identities,
\begin{equation}
  \begin{aligned}
    \frac{1}{m^2}\sum_{x,y=1}^{m}G_{xy}^2
    &=\frac{1}{16},\\
    \frac{1}{m^2}\sum_{x,y=1}^{m}G_{xy}^4
    &=\frac{3}{16\cdot18}
    =\frac{1}{96}.
  \end{aligned}
\end{equation}
These even moments are unchanged by retaining one representative from each
antipodal pair.

Let $L_{\mathrm{heur}}(M)$ denote the largest binary value found by the
heuristic. Table~\ref{tab:lattice-shell-bounds} reports
\begin{equation}
  \frac{\widehat Q_{16}(M)}{L_{\mathrm{heur}}(M)}
  \approx 1.682047299161634.
  \label{eq:BW16-candidate-ratio}
\end{equation}
Since $L_{\mathrm{heur}}(M)\leq L(M)$, this ratio alone does not establish a
dimension witness. A sufficient condition is given below.

\medskip \noindent \textbf{Proposition.} If the matrix $M$ defined above
satisfies
\begin{equation}
  L(M)<\frac{\widehat Q_{16}(M)}{1.649646},
  \label{eq:dimension-witness-certification-target}
\end{equation}
then the correlations $E_{xy}=\langle v_x,v_y\rangle$ certify Schmidt number
at least five. The Schmidt number of a mixed state is the smallest integer $r$
for which it admits a convex decomposition into pure states of Schmidt rank at
most $r$~\cite{Terhal2000}. In particular, every quantum realization of these
correlations has smaller local Hilbert-space dimension at least five. The
correlations $E_{xy}=\langle v_x,v_y\rangle$ can be realized using a maximally
entangled state of local dimension $256$; under
condition~\eqref{eq:dimension-witness-certification-target}, they violate the
dimension witness.

\smallskip \noindent \textit{Proof.} By
Eq.~\eqref{eq:vertesi-pal-dimension-bound47} and Krivine's upper bound on
$K_G(7)$~\cite{Krivine1979}, listed in Table~\ref{tab:comparison-bounds},
\begin{equation}
  \begin{aligned}
    W_4(M)
    &\leq Q_7(M)\\
    &\leq K_G(7)L(M)\\
    &\leq 1.649646\,L(M).
  \end{aligned}
  \label{eq:four-dimensional-quantum-upper-bound}
\end{equation}
Every pure state of Schmidt rank at most four satisfies this bound, since its
local supports have dimension at most four. By convexity, the same is true for
every mixed state of Schmidt number at most four, regardless of its ambient
local dimensions. Thus all such states, and in particular all realizations
whose smaller local dimension is at most four, satisfy
\begin{equation}
  \sum_{x,y}M_{xy}E_{xy}\leq 1.649646\,L(M).
\end{equation}

As explained in the preceding subsection, Tsirelson's construction realizes
the correlators $E_{xy}=\langle v_x,v_y\rangle$, where
$v_x,v_y\in\mathbb{R}^{16}$ are unit vectors, using binary observables and a
maximally entangled state of local dimension $D'=2^{\lfloor16/2\rfloor}=256$
\cite{Tsirelson1980,Tsirelson1987}. For completeness, choose Hermitian
matrices $\Gamma_1,\ldots,\Gamma_{16}$ on $\mathbb C^{256}$ satisfying
\begin{equation}
  \Gamma_i\Gamma_j+\Gamma_j\Gamma_i
  =2\delta_{ij}\mathbb I_{256},
\end{equation}
and set
\begin{equation}
  \begin{aligned}
    A_x&=\sum_{i=1}^{16}(v_x)_i\Gamma_i,\\
    B_y&=\sum_{i=1}^{16}(v_y)_i\Gamma_i^{\mathsf T}.
  \end{aligned}
  \label{eq:BW16-clifford-observables}
\end{equation}
These observables satisfy $A_x^2=B_y^2=\mathbb I_{256}$. On the state
\begin{equation}
  \ket{\Phi_{256}}
  =\frac{1}{16}\sum_{j=1}^{256}\ket{j}\ket{j},
\end{equation}
they give
\begin{equation}
  \begin{aligned}
    \bra{\Phi_{256}}A_x\otimes B_y\ket{\Phi_{256}}
    &=\frac{\operatorname{Tr}(A_xB_y^{\mathsf T})}{256}\\
    &=\langle v_x,v_y\rangle.
  \end{aligned}
\end{equation}
The last equality follows from
$\operatorname{Tr}(\Gamma_i\Gamma_j)=256\delta_{ij}$. The resulting witness
value is $\widehat Q_{16}(M)$. Under
condition~\eqref{eq:dimension-witness-certification-target}, it exceeds the
upper bound in Eq.~\eqref{eq:four-dimensional-quantum-upper-bound}, proving
the claim. \hfill$\square$

Here $7$ is the dimension of the real vectors used to bound the quantum value,
whereas $4$ is the excluded local Hilbert-space dimension. The construction
shows that local dimension $256$ is sufficient to realize the prescribed
correlations; it does not show that this dimension is necessary.

Indeed, the eighth nonempty shell of the Coxeter--Todd lattice $K_{12}$, of
squared norm $18$, already provides a candidate witness with $m=510\,048$
binary measurement settings per party and a realization in local dimension
$2^6=64$. Its heuristic ratio
\begin{equation}
  \frac{\widehat Q_{12}(M)}{L_{\mathrm{heur}}(M)}
  \approx1.658738
\end{equation}
exceeds the bound $1.649646$ for $K_G(7)$ (see
Table~\ref{tab:lattice-shell-bounds}). Thus this smaller construction would
also certify local dimension at least five for each party, provided that
\begin{equation}
  L(M)<\frac{\widehat Q_{12}(M)}{1.649646}.
\end{equation}

The dimension-witness conclusion for the explicit constructions above remains
conditional on a sufficiently strong upper bound on the true binary optimum
$L(M)$. The heuristic ratios provide numerical evidence for the required
separation, but do not prove it.

The existence of a witness certifying local dimension at least five for each
party already follows from the Davie--Reeds bound~\cite{Davie1984,Reeds1991}:
\begin{equation}
  \begin{aligned}
    K_G
    &\geq1.676956674\ldots\\
    &>1.649646
    \geq K_G(7).
  \end{aligned}
\end{equation}
Their construction uses a continuum of measurement settings, whereas our
candidate witnesses are explicitly finite. The strict gap also guarantees the
existence of a finite-setting witness, since $K_G$ is defined as a supremum
over finite matrices.

\section{Discussion}
\label{sec:conclusion}

We have introduced a rotationally invariant cubic kernel that yields candidate
lower bounds on $\KG(d)$ for every $d\geq3$. A stability analysis identifies
the threshold
\begin{equation}
  c_*(d)=\frac{(d+1)(d+3)}{3(d+5)},
\end{equation}
at which the hemispherical binary strategy becomes marginal against a
degree-three perturbation. Large-scale see-saw calculations for discrete
approximations in dimensions up to $d=24$ provide numerical support for the
continuous predictions.

For the fifth shell of the Barnes--Wall lattice $BW_{16}$, we obtain the
finite-setting candidate ratio
\begin{equation}
  \frac{\widehat Q_{16}(M)}{L_{\mathrm{heur}}(M)}
  \approx1.68204729.
\end{equation}
This exceeds the classical Davie--Reeds lower bound
$\KG\geq1.676956674\ldots\approx1.676957$ for the infinite-order Grothendieck
constant. It also provides a candidate witness of Schmidt number at least
five, subject to the certification condition discussed in
Sec.~\ref{sec:physics}. For the continuous problem in dimension $24$, our
prediction is
\begin{equation}
  \KG^{\mathrm{cub}}(24)
  =1.7185395336\ldots
  \approx1.718540.
\end{equation}
If certified, this would establish the same lower bound for $\KG(24)$ and
hence for $\KG$, improving on the rigorous bound
$\KG\geq6\pi/11\approx1.713596$ of Ref.~\cite{Saha2026}. It would also yield a
dimension witness certifying Schmidt number at least five and thus local
Hilbert-space dimension at least five for each party. These conclusions remain
conditional, since the binary optimization has not yet been certified.

The central open problem is to prove the global hemispherical-optimality
conjecture
\begin{equation}
  L(M_c)=L_{\mathrm{HS}}(d,c),
  \qquad 0\leq c\leq c_*(d).
  \label{eq:open-hemispherical-problem}
\end{equation}
The local stability analysis does not rule out a competing binary strategy
with a larger value. However, the full conjecture reduces to proving global
hemispherical optimality at $c=c_*(d)$: convexity of $L(M_c)$, affinity of the
hemispherical value, and their equality at $c=0$ then imply equality
throughout the interval, as shown in
Sec.~\ref{subsec:conjectured-cubic-bounds}. This would prove, for every finite
$d\geq3$,
\begin{equation}
  \KG(d)\geq\KG^{\mathrm{cub}}(d)
  =
  \frac{\pi\Gamma\!\left(\frac{d+1}{2}\right)^2}
  {d\Gamma\!\left(\frac d2\right)^2}
  \frac{3(d+1)(3d+7)}
  {4(d+2)(2d+3)},
  \label{eq:open-finite-bound}
\end{equation}
and, by taking $d\to\infty$,
\begin{equation}
  \KG\geq\frac{9\pi}{16}\approx1.767146.
\end{equation}

A useful intermediate goal is to obtain rigorous upper bounds on the binary
optima of the finite matrices, particularly the $BW_{16}$ instance.
Determining these optima exactly would establish the reported candidate ratios.
For the dimension-witness application, a weaker upper bound already suffices,
as specified in Eq.~\eqref{eq:dimension-witness-certification-target}.

Beyond proving these lower bounds, one may ask whether the cubic prediction
gives the exact value of $\KG(d)$ in some higher dimension, or even whether
\begin{equation}
  \KG=\frac{9\pi}{16}.
\end{equation}
Exactness is already ruled out in dimension three:
$\KG^{\mathrm{cub}}(3)=64/45\approx1.422222$, whereas
$\KG(3)\geq1.43670$~\cite{Designolle2026}. This does not settle the question
in higher dimensions or in the infinite-dimensional limit.

Proving that these bounds are exact would require matching upper bounds.
Alternatively, better lower bounds could be sought using higher-degree kernels
or other matrix constructions.

\begin{acknowledgments}
We thank S\'ebastien Designolle for fruitful discussions.

GPT-5.6 Sol and GPT-6 Astra assisted us in preparing this manuscript. Our work
on cubic kernels began in 2022 and was recently accelerated by these models.
Their most notable contribution was helping us derive the expression for the
critical parameter $c_*$ in Eq.~\eqref{eq:cstar-gen}.

We acknowledge support from the European Union through the CHIST-ERA MoDIC
project and from the National Research, Development and Innovation Office
(NKFIH) under Grant Nos.~2023-1.2.1-ERA\_NET-2023-00009 and K145927.
T.V. acknowledges support from the NKFIH's ``Frontline'' Research
Excellence Program under Grant No.~KKP133827.
\end{acknowledgments}

\appendix

\section{Derivation of the critical parameter in arbitrary dimension}
\label{app:crit}

We extend the boundary-deformation calculation of Sec.~\ref{sec:cstar-d3} to
arbitrary $d\geq3$. The argument applies to any transverse harmonic cubic,
including the explicit polynomial used in the main text.

Write
\begin{equation}
  \begin{gathered}
    x=\left(\sqrt{1-u^2}\,w,u\right),\\
    w\in S^{d-2},
    \qquad
    u=x_d.
  \end{gathered}
\end{equation}
With the convention that $\sigma_d$ is normalized measure on $S^{d-1}$, the
spherical measure decomposes as
\begin{equation}
  \mathrm{d}\sigma_d(x)
  =
  \rho_d(u)\,\mathrm{d}u\,
  \mathrm{d}\sigma_{d-1}(w),
\end{equation}
where
\begin{equation}
  \begin{aligned}
    \rho_d(u)
    &=
    \rho_0(1-u^2)^{(d-3)/2},\\
    \rho_0
    &=
    \frac{\Gamma(d/2)}
    {\sqrt{\pi}\Gamma((d-1)/2)}.
  \end{aligned}
  \label{eq:rho-general}
\end{equation}
The absolute moment introduced in Eq.~\eqref{eq:mu-d-Davie} satisfies
\begin{equation}
  \begin{aligned}
    \mu_d
    &=
    2\rho_0\int_0^1
    u(1-u^2)^{(d-3)/2}\,\mathrm{d}u\\a
    &=
    \frac{2\rho_0}{d-1}.
  \end{aligned}
  \label{eq:mu-rho-relation}
\end{equation}

Let $h(w)$ be the restriction to $S^{d-2}$ of a nonzero real homogeneous
harmonic cubic in the transverse coordinates. Its value on $S^{d-1}$ can be
written as
\begin{equation}
  H(x)=(1-u^2)^{3/2}h(w).
\end{equation}
We use the normalization
\begin{equation}
  \|h\|_2^2
  :=
  \int_{S^{d-2}}
  h(w)^2\,\mathrm{d}\sigma_{d-1}(w)
  >0.
  \label{eq:h-norm-general}
\end{equation}
Consider the opposite deformations
\begin{align}
  a_s(x)&=\operatorname{sgn}\!\left[u+sH(x)\right],
  \notag\\
  b_s(x)&=\operatorname{sgn}\!\left[u-sH(x)\right].
\end{align}
Their sign boundaries satisfy
\begin{align}
  u_a(w)&=-s h(w)+O(s^3),
  \notag\\
  u_b(w)&=+s h(w)+O(s^3).
  \label{eq:boundaries-general}
\end{align}
These expansions follow from the implicit boundary equations and the even
dependence of $(1-u^2)^{3/2}$ on $u$.

As in the three-dimensional calculation, set $\delta a=a_s-a_0$ and $\delta
b=b_s-b_0$, with $a_0(x)=b_0(x)=\operatorname{sgn}(u)$. We evaluate separately
the change produced by each boundary shift and the mixed contribution.

\subsection{Contribution from each boundary shift}

The field generated by a hemispherical strategy is
\begin{equation}
  \begin{aligned}
    &F_c(u)\\
    &\quad:=
    \int_{S^{d-1}}
    k_c(\langle x,y\rangle)
    \operatorname{sgn}(y_d)\,
    \mathrm{d}\sigma_d(y)\\
    &\quad=
    \mu_d u
    \left[
    1-\frac{c}{d+1}(3-u^2)
    \right].
  \end{aligned}
  \label{eq:hemisphere-local-field}
\end{equation}
To obtain this expression, expand the cubic term and use
\begin{equation}
  \begin{aligned}
    \int |y_d|^3\,\mathrm{d}\sigma_d(y)
    &=
    \frac{2\mu_d}{d+1},\\
    \int |y_d|y_j^2\,\mathrm{d}\sigma_d(y)
    &=
    \frac{\mu_d}{d+1},
    \qquad j<d.
  \end{aligned}
\end{equation}
The slope of the field at the equator is therefore
\begin{equation}
  F_c'(0)
  =
  \mu_d\left(1-\frac{3c}{d+1}\right).
  \label{eq:hemisphere-field-slope}
\end{equation}

Shifting Alice's boundary gives
\begin{equation}
  \begin{aligned}
    &\int F_c(u)\delta a(x)\,\mathrm{d}\sigma_d(x)\\
    &\quad=
    -2\int_{S^{d-2}}
    \int_0^{u_a(w)}
    \rho_d(u)F_c(u)\,\mathrm{d}u\\
    &\qquad\qquad\times
    \mathrm{d}\sigma_{d-1}(w)\\
    &\quad=
    -\rho_0F_c'(0)\|h\|_2^2s^2+O(s^4).
  \end{aligned}
  \label{eq:single-boundary-general}
\end{equation}
Here $\rho_d(u)$ is even and $F_c(u)$ is odd, so the leading term in the inner
integral is proportional to $u_a(w)^2$. Bob's contribution is the same.
Together, the two terms give
\begin{equation}
  \begin{aligned}
    &-2\mu_d\rho_0
    \left(1-\frac{3c}{d+1}\right)
    \|h\|_2^2s^2\\
    &\qquad+O(s^4).
  \end{aligned}
  \label{eq:separate-boundaries-general}
\end{equation}

\subsection{Interaction between the two deformations}

The integrated change in Alice's sign function at fixed $w$ is
\begin{equation}
  \begin{aligned}
    &\int_{-1}^{1}
    \delta a(u,w)\rho_d(u)\,\mathrm{d}u\\
    &\quad=
    -2\int_0^{u_a(w)}\rho_d(u)\,\mathrm{d}u\\
    &\quad=
    2\rho_0s h(w)+O(s^3),
  \end{aligned}
\end{equation}
whereas Bob's change has the opposite sign. Moreover, the scalar product has
no term linear in either longitudinal coordinate at the equator. Replacing the
kernel by its equatorial value therefore changes the mixed contribution only
at order $s^4$. We obtain
\begin{equation}
  \begin{aligned}
    &\int\!\!\int
    k_c(\langle x,y\rangle)\delta a(x)\delta b(y)\\
    &\qquad\times
    \mathrm{d}\sigma_d(x)\mathrm{d}\sigma_d(y)\\
    &\quad=
    -4\rho_0^2s^2
    \int\!\!\int
    k_c(\langle w,z\rangle)h(w)h(z)\\
    &\qquad\times
    \mathrm{d}\sigma_{d-1}(w)\mathrm{d}\sigma_{d-1}(z)
    +O(s^4).
  \end{aligned}
  \label{eq:mixed-boundary-general}
\end{equation}

The angular integrals are determined by two identities:
\begin{equation}
  \begin{aligned}
    &\int_{S^{d-2}}
    \langle w,z\rangle h(z)\,
    \mathrm{d}\sigma_{d-1}(z)
    =0,\\
    &\int_{S^{d-2}}
    \langle w,z\rangle^3 h(z)\,
    \mathrm{d}\sigma_{d-1}(z)\\
    &\quad=
    \frac{6h(w)}
    {(d-1)(d+1)(d+3)}.
  \end{aligned}
  \label{eq:cubic-equatorial-eigenvalue}
\end{equation}
The first follows from orthogonality of degree-one and degree-three spherical
harmonics.

For completeness, the second identity can be obtained directly from spherical
moments. Set $n=d-1$ and write
\begin{equation}
  h(z)=\sum_{i,j,k=1}^{n}T_{ijk}z_i z_j z_k,
\end{equation}
where $T$ is symmetric and traceless:
\begin{equation}
  T_{ijk}=T_{(ijk)},
  \qquad
  \sum_{i=1}^{n}T_{iik}=0.
\end{equation}
This is the tensor representation of a homogeneous harmonic
cubic~\cite{DaiXu2013,AtkinsonHan2012}. The sixth-order spherical moment
formula has denominator $n(n+2)(n+4)$ and sums over the $15$ pairings of its
six indices~\cite{Folland2001}. Tracelessness eliminates all pairings that
join two indices of $T$. The remaining six pairings each contribute $h(w)$,
giving Eq.~\eqref{eq:cubic-equatorial-eigenvalue}. The same argument gives
\begin{equation}
  \begin{aligned}
    \|h\|_2^2
    &=
    \frac{6\|T\|_{\mathrm F}^2}
    {n(n+2)(n+4)},\\
    \|T\|_{\mathrm F}^2
    &=
    \sum_{i,j,k=1}^{n}T_{ijk}^2.
  \end{aligned}
  \label{eq:spherical-moments-general}
\end{equation}

It follows from Eq.~\eqref{eq:cubic-equatorial-eigenvalue} that
\begin{equation}
  \begin{aligned}
    &\int\!\!\int
    k_c(\langle w,z\rangle)h(w)h(z)\\
    &\qquad\times
    \mathrm{d}\sigma_{d-1}(w)\mathrm{d}\sigma_{d-1}(z)\\
    &\quad=
    -\frac{6c}{(d-1)(d+1)(d+3)}\|h\|_2^2.
  \end{aligned}
\end{equation}
Thus the mixed contribution in Eq.~\eqref{eq:mixed-boundary-general} is
\begin{equation}
  \frac{24c\rho_0^2}
  {(d-1)(d+1)(d+3)}
  \|h\|_2^2s^2+O(s^4).
  \label{eq:mixed-contribution-general}
\end{equation}

\subsection{The quadratic variation}

Combining Eqs.~\eqref{eq:separate-boundaries-general} and
\eqref{eq:mixed-contribution-general}, and using $\rho_0=(d-1)\mu_d/2$, gives
\begin{equation}
  \begin{aligned}
    &\mathcal{L}_c(a_s,b_s)
    =
    L_{\mathrm{HS}}(d,c)\\
    &\quad+
    \mu_d\rho_0\|h\|_2^2
    \left[
    -2+\frac{6c(d+5)}{(d+1)(d+3)}
    \right]s^2\\
    &\quad+O(s^4).
  \end{aligned}
  \label{eq:second-variation-general-app}
\end{equation}
The expansion is even in $s$, since changing its sign exchanges the two
parties. Its quadratic coefficient vanishes when
\begin{equation}
  \frac{3c(d+5)}{(d+1)(d+3)}=1,
\end{equation}
which yields the critical parameter in Eq.~\eqref{eq:cstar-general}.

For the explicit polynomial $H(x)=x_1^3-3x_1x_2^2$ used in the main text, the
nonzero tensor components are
\begin{equation}
  \begin{gathered}
    T_{111}=1,\\
    T_{122}=T_{212}=T_{221}=-1.
  \end{gathered}
\end{equation}
Hence $\|T\|_{\mathrm F}^2=4$, and
\begin{equation}
  \|h\|_2^2
  =
  \frac{24}{(d-1)(d+1)(d+3)}.
\end{equation}
Substituting this expression into Eq.~\eqref{eq:second-variation-general-app}
gives Eq.~\eqref{eq:second-variation-general}. In particular, $d=3$ gives
$\mu_3=\rho_0=1/2$ and $\|h\|_2^2=1/2$, recovering the correction $(c-1)s^2/4$
derived in the main text.

For any nonzero transverse harmonic cubic, $\mu_d\rho_0\|h\|_2^2$ is positive.
All such deformations therefore become marginal at the same value $c_*(d)$.
For fixed $\|h\|_2$, their quadratic corrections coincide; differences between
their shapes can first enter at fourth or higher order.

\section{Large-scale see-saw optimization}
\label{app:seesaw}

For a fixed coefficient matrix $M_c$ defined by Eq.~\eqref{eq:discrete-Mc}, we
seek the classical binary optimum
\begin{align}
  L(M_c)
  &=\max_{a,b\in\{-1,+1\}^{m}}a^{\mathsf T}M_cb
  \notag\\
  &=\max_{b\in\{-1,+1\}^{m}}
  \sum_{x=1}^{m}|(M_cb)_x|.
\end{align}
Our numerical method is heuristic: it searches for good binary strategies
without guaranteeing the global optimum. The strategy returned has a value
$L_{\mathrm{heur}}(M_c)\leq L(M_c)$. Consequently, the ratio $\widehat
Q_d(M_c)/L_{\mathrm{heur}}(M_c)$ provides a candidate lower bound on $\KG(d)$,
whose certification requires a suitable upper bound on the true binary
optimum.

For fixed $b$, the optimal choice of Alice's signs is
\begin{align}
  a_x=\sgn\!\left((M_cb)_x\right),
  \label{eq:best-response-a}
\end{align}
and, for fixed $a$, Bob's optimal signs are
\begin{align}
  b_y=\sgn\!\left((M_c^{\mathsf T}a)_y\right).
  \label{eq:best-response-b}
\end{align}
We take $\sgn(0)=1$. Alternating these updates gives the see-saw
method~\cite{Araujo2020,Marton2024}, based on the same
alternating-optimization principle widely used in quantum correlation
problems~\cite{Werner2001,Liang2009,PalVertesi2010,PratColomerEtAl2022}. Since
$M_c$ is symmetric, Bob's update can equivalently be written as
$b_y=\sgn\!\left((M_ca)_y\right)$. In exact arithmetic, each update maximizes
the objective over one party's signs and cannot decrease its value. At a fixed
point, neither party can improve the value while the other party's signs
remain fixed. This does not imply global optimality.

We therefore repeat the search from many starting strategies \cite{Marti2003}.
Several strategies are processed together in a batch, so their matrix products
can be evaluated efficiently in a single operation. Each strategy nevertheless
undergoes its own sequence of updates. The implementation cycles through six
types of initial sign vectors:
\begin{enumerate}
  \item Independent random signs, with equal probabilities
  for $+1$ and $-1$.

  \item Random hemispheres,
  $b_y=\sgn\langle v_y,g\rangle$, where $g$ is a vector of independent
  standard Gaussian variables.

  \item Signs of random polynomials containing linear and
  cubic terms in the coordinates of $v_y$, with Gaussian coefficients.

  \item One-step responses $b=\sgn(M_cr)$ to independent
  random sign vectors $r$.

  \item Coordinate hemispheres,
  $b_y=\sgn((v_y)_j)$, for selected coordinate directions $j$.

  \item Random hemispheres in which each sign is independently
  flipped with probability $0.05$.
\end{enumerate}
These choices explore both unstructured strategies and strategies reflecting
the geometry of the vectors.

Before applying the noiseless see-saw updates, we optionally apply a short
stage inspired by annealing~\cite{Kirkpatrick1983}. Gaussian noise is added to
the quantities entering the sign functions, and its relative strength is
gradually reduced. These perturbations help the search explore other
strategies and may temporarily decrease the objective.

The noise is then removed, and the ordinary see-saw updates are applied to
refine each strategy. The implementation stops a batch when all right-hand
sign vectors remain unchanged, when all improvements fall below a prescribed
tolerance, or when the iteration limit is reached. The tolerance test is
applied only after at least six complete see-saw sweeps. Stopping because the
signs are unchanged gives a fixed point; the other stopping rules do not
guarantee this. In every case, Alice's signs are recomputed from the final
$b$, giving the consistent score
\begin{equation}
  a^{\mathsf T}M_cb
  =\sum_{x=1}^{m}|(M_cb)_x|.
\end{equation}
We retain the best score among the refined strategies.

After the initial searches, we generate additional starting points by
perturbing copies of the best strategy found so far. Each sign is
independently flipped with a probability that decreases geometrically between
trials, with a minimum probability of $1/m$. If no sign is selected, one
randomly chosen sign is flipped. Each perturbed strategy undergoes the same
optional noisy stage and noiseless refinement. The best stored strategy is
replaced only when an improvement is found. This repeated perturbation and
refinement follows the idea of iterated local search
\cite{LourencoMartinStutzle2003}. The complete procedure is summarized in
Fig.~\ref{alg:discrete-seesaw}.

\begin{figure}[t]
  \hrule \smallskip \textbf{See-saw algorithm for estimating $L(M_c)$}
  \begin{enumerate}
    \item Prepare the cubic monomials and their multiplicities
    needed in Eq.~\eqref{eq:implicit-M-product}. Store their values if memory
    permits; otherwise compute them in blocks.

    \item Generate a batch of starting sign vectors using
    the six choices described above.

    \item Optionally apply noisy alternating updates,
    gradually reducing the relative noise strength.

    \item Apply the noiseless updates
    $a\leftarrow\sgn(M_cb)$ and $b\leftarrow\sgn(M_c^{\mathsf T}a)$ until a
    stopping criterion is met.

    \item Recompute $a=\sgn(M_cb)$ for each final $b$,
    evaluate $\sum_x|(M_cb)_x|$, and update the best stored strategy.

    \item Repeat steps 2--5 until the prescribed number
    of initial strategies has been processed.

    \item Generate batches of perturbed copies of the best
    strategy by randomly flipping signs. Apply steps 3--5 to each batch until
    the prescribed number of perturbations has been processed.

    \item Return the best strategy and its value
    $L_{\mathrm{heur}}(M_c)$.
  \end{enumerate}
  \smallskip \hrule
  \caption{Heuristic search for the classical binary optimum.
    All products with $M_c$ are evaluated using
    Eq.~\eqref{eq:implicit-M-product}, without forming the full coefficient
    matrix.}
  \label{alg:discrete-seesaw}
\end{figure}

Our largest instances contain several million settings per party. Storing the
full coefficient matrix would require $m^2$ entries, so both the binary search
and the evaluation of the explicit vector value $\widehat Q_d(M_c)$ must avoid
forming this matrix. The cubic form of Eq.~\eqref{eq:discrete-Mc} makes this
possible.

Let $V\in\mathbb{R}^{m\times d}$ have rows $v_x^{\mathsf T}$, and write
$G=VV^{\mathsf T}$ for the Gram matrix. We collect all distinct cubic
coordinate products into a matrix $\Phi_3\in\mathbb{R}^{m\times D_3}$, where
$D_3=\binom{d+2}{3}$:
\begin{align}
  (\Phi_3)_{x,(i,j,k)}
  =(v_x)_i(v_x)_j(v_x)_k,
  \qquad 1\leq i\leq j\leq k\leq d.
\end{align}
Let $\Lambda_3$ be diagonal, with entries equal to $1$ when all three indices
coincide, $3$ when exactly two coincide, and $6$ when all three are distinct.
These multiplicities account for the different orderings of the indices in the
expansion of $\langle v_x,v_y\rangle^3$. Thus
\begin{align}
  G_{xy}^3
  =(\Phi_3\Lambda_3\Phi_3^{\mathsf T})_{xy},
\end{align}
and
\begin{align}
  M_cz
  =V(V^{\mathsf T}z)
  -c\Phi_3\Lambda_3(\Phi_3^{\mathsf T}z).
  \label{eq:implicit-M-product}
\end{align}
The same expression applies to several sign vectors at once by arranging them
as columns of $z$.

For fixed $d$ and batch size, the storage and the work per matrix product grow
linearly with $m$. Storing $\Phi_3$ requires $mD_3$ entries, which can still
be substantial because $D_3$ grows cubically with $d$. When this exceeds the
chosen memory allowance, the implementation generates the rows of $\Phi_3$ in
blocks. A first pass accumulates $\Phi_3^{\mathsf T}z$, and a second pass
evaluates the cubic contribution to $M_cz$. This reduces memory use at the
cost of recomputing the cubic monomials. All formulas here retain the diagonal
of $M_c$, which equals $1-c$ for unit vectors.

The explicit vector value in Eq.~\eqref{eq:discrete-Qhat} requires no
optimization:
\begin{equation}
  \widehat Q_d(M_c)
  =\sum_{x,y=1}^{m}G_{xy}^2
  -c\sum_{x,y=1}^{m}G_{xy}^4.
\end{equation}
Both sums can be evaluated without constructing $G$. For the second-order
term,
\begin{align}
  \sum_{x,y=1}^{m}G_{xy}^{2}
  &=\lVert V^{\mathsf T}V\rVert_{\mathrm F}^{2},
  \label{eq:second-moment-fast}
\end{align}
where $\|\cdot\|_{\mathrm F}$ denotes the Frobenius norm. For the fourth-order
term, expanding $\langle v_x,v_y\rangle^4$ gives
\begin{align}
  \sum_{x,y=1}^{m}G_{xy}^{4}
  &=\sum_{|\alpha|=4}\binom{4}{\alpha}
  \left(\sum_{x=1}^{m}v_x^\alpha\right)^2.
  \label{eq:fourth-moment-fast}
\end{align}
Here $\alpha=(\alpha_1,\ldots,\alpha_d)$ is a tuple of nonnegative integers
with $|\alpha|=\sum_j\alpha_j=4$, and
\begin{equation}
  v_x^\alpha=\prod_{j=1}^{d}(v_x)_j^{\alpha_j},
  \qquad
  \binom{4}{\alpha}
  =\frac{4!}{\alpha_1!\cdots\alpha_d!}.
\end{equation}
In practice, we sum each distinct fourth-degree coordinate product over the
vectors, square the resulting sum, and multiply by its multinomial
multiplicity. Only $\binom{d+3}{4}$ such sums are needed, and they can be
accumulated in blocks. Together with Eq.~\eqref{eq:implicit-M-product}, these
identities avoid quadratic storage in $m$ and make the large instances
considered here accessible.

\section{Construction of the generating vectors}
\label{app:const}

This appendix describes the sets of $m$ unit vectors
$v_1,\ldots,v_m\in\mathbb{R}^d$ used to construct the coefficient matrices in
Eq.~\eqref{eq:discrete-Mc}. Following the notation of
Eq.~\eqref{eq:discrete-Gram}, we denote the Gram matrix of each set by
$G\in\mathbb{R}^{m\times m}$, with entries
\begin{equation}
  G_{xy}=\langle v_x,v_y\rangle.
\end{equation}

We use spiral constructions in dimensions $3$ and $4$, Sobol-based
constructions in dimensions $6$ and $8$, and lattice shells in dimensions $8$,
$12$, $16$, and $24$. The MATLAB routines for generating the vector sets
accompany the arXiv submission.

The shell populations are commonly listed through the lattice's theta series,
\begin{equation}
  \Theta_\Lambda(q)
  =1+\sum_{n=1}^{\infty}N_nq^n,
  \qquad
  N_n=|\mathcal{S}_n(\Lambda)|.
  \label{eq:appendix-theta-series}
\end{equation}
Here the coefficient $N_n$ simply counts the vectors of squared length $2n$,
including both members of every antipodal pair. These tabulated counts
therefore give $m=N_n/2$. When a lattice has no vectors of squared length $2$,
its first nonempty shell has index $n=2$.

\subsection{Spiral constructions, \texorpdfstring{$d=3$}{d=3} and \texorpdfstring{$d=4$}{d=4}}
\label{app:const_d34}

In dimensions $3$ and $4$, we use spiral constructions to generate $m$ unit
vectors that approximate uniform spherical sampling. In both cases, the
vectors are indexed by $i=1,\ldots,m$.

For $d=3$, we use a spherical Fibonacci spiral~\cite{Gonzalez2010}. Define
\begin{align*}
  z_i&=1-\frac{2i-1}{m},
  &
  r_i&=\sqrt{1-z_i^2},\\
  \varphi_i&=(i-1)\pi(3-\sqrt{5}).
\end{align*}
The generating vectors are
\begin{equation}
  v_i=
  \bigl(
  r_i\cos\varphi_i,\,
  r_i\sin\varphi_i,\,
  z_i
  \bigr).
  \label{eq:fibonacci-S2}
\end{equation}
Each vector has unit norm in $\mathbb{R}^3$. The equally spaced heights place
one point in each equal-area band, while the golden-angle increment
distributes the points around the sphere. This construction is implemented in
\texttt{fibonacci\_sphere\_D3.m}.

For $d=4$, we use the super-Fibonacci construction~\cite{Alexa2022}. Set
$s_i=i-\tfrac12$ and define
\begin{align*}
  r_i&=\sqrt{\frac{s_i}{m}},
  &
  R_i&=\sqrt{1-\frac{s_i}{m}},\\
  \alpha_i&=\frac{2\pi s_i}{\sqrt{2}},
  &
  \beta_i&=\frac{2\pi s_i}{\psi},
\end{align*}
where $\psi=1.533751168755204\ldots$ is the constant recommended in
Ref.~\cite{Alexa2022}. The generating vectors are
\begin{equation}
  v_i=
  \bigl(
  r_i\sin\alpha_i,\,
  r_i\cos\alpha_i,\,
  R_i\sin\beta_i,\,
  R_i\cos\beta_i
  \bigr).
  \label{eq:super-fibonacci-S3}
\end{equation}
Since $r_i^2+R_i^2=1$, each vector has unit norm in $\mathbb{R}^4$. This
construction is implemented in \texttt{super\_fibonacci\_sphere\_D4.m}.

\subsection{Sobol-based constructions, \texorpdfstring{$d=6$}{d=6} and \texorpdfstring{$d=8$}{d=8}}
\label{app:const_d68}

For $d=6$ and $d=8$, we generate unit vectors from Sobol
sequences~\cite{Sobol1967}, using \texttt{sobol\_sphere\_D6.m} and
\texttt{sobol\_sphere\_D8.m}, respectively. We omit the first two points,
$(0,\ldots,0)$ and $(1/2,\ldots,1/2)$: the first gives divergent inverse
normal values, while the second gives a zero vector that cannot be normalized.

Let $u_i=(u_{i1},\ldots,u_{id})\in(0,1)^d$ denote the retained points, indexed
by $i=1,\ldots,m$, and let $\Phi$ be the cumulative distribution function of
the standard normal distribution. We define
\begin{align}
  g_i&=
  \bigl(
  \Phi^{-1}(u_{i1}),\ldots,\Phi^{-1}(u_{id})
  \bigr),
  \notag\\
  v_i&=\frac{g_i}{\|g_i\|_2}.
  \label{eq:sobol-sphere-vectors}
\end{align}
For independent uniformly distributed inputs, this transformation produces an
isotropic Gaussian vector, whose normalization gives a uniformly distributed
direction~\cite{Muller1959}. Applying the same transformation to Sobol points
gives a deterministic approximation to uniform spherical sampling in
$\mathbb{R}^d$. In our calculations, $m=2^k$ denotes the number of vectors
retained after omitting the two initial points.

For the lattice constructions, a shell consists of all lattice vectors at a
fixed distance from the origin. We use the standard normalizations in which
squared lengths are even integers and define
\begin{equation}
  \mathcal{S}_n(\Lambda)
  :=
  \{w\in\Lambda:\|w\|^2=2n\},
  \qquad n\geq1.
  \label{eq:appendix-lattice-shell}
\end{equation}
Every vector $w$ in a shell is accompanied by $-w$. We retain one vector from
each such pair and normalize it:
\begin{equation}
  v=\frac{w}{\sqrt{2n}},
  \qquad
  m=\frac{|\mathcal{S}_n(\Lambda)|}{2}.
\end{equation}
Thus $m$ always denotes the number of retained unit vectors.

\subsection{Shells of the \texorpdfstring{$E_8$}{E8} lattice, \texorpdfstring{$d=8$}{d=8}}
\label{app:const_d8}

Our second construction in dimension $8$ uses shells of the $E_8$
lattice~\cite{Conway1999}. In its standard normalization,
\begin{equation}
  E_8=
  \left\{
  w\in\mathbb{Z}^8\cup
  \left(\mathbb{Z}+\tfrac12\right)^8:
  \sum_{j=1}^{8}w_j\in2\mathbb{Z}
  \right\}.
  \label{eq:E8-lattice}
\end{equation}
For enumeration, we use the integer coordinates $a=2w$. Their entries are
either all even or all odd, and their sum is divisible by $4$. The shell of
index $n$ is obtained by imposing
\begin{equation}
  \sum_{j=1}^{8}a_j^2=8n.
  \label{eq:E8-shell-construction}
\end{equation}
After retaining one representative from each antipodal pair, the generating
vectors are $v=a/\sqrt{8n}$.

The number of retained vectors is
\begin{equation}
  m=120\sigma_3(n),
  \qquad
  \sigma_3(n)=\sum_{r\mid n}r^3,
  \label{eq:E8-shell-size}
\end{equation}
where the sum runs over the positive divisors of $n$. In particular, the
shortest vectors have squared norm $2$ and give $m=120$. The full shell
populations are tabulated in OEIS~A004009~\cite{OEISA004009}.

\subsection{Shells of the Coxeter--Todd lattice, \texorpdfstring{$d=12$}{d=12}}
\label{app:const_d12}

For $d=12$, we use shells of the Coxeter--Todd lattice
$K_{12}$~\cite{Coxeter1953,Conway1983}. In the normalization used here, its
shortest nonzero vectors have squared norm $4$. They form the shell of index
$n=2$ and give $m=378$ unit vectors after retaining one representative from
each antipodal pair and normalizing.

Higher shells are treated in the same way, with their populations listed in
OEIS~A004010~\cite{OEISA004010}. In particular, the eighth nonempty shell has
index $n=9$, squared norm $18$, and yields $m=510\,048$ unit vectors. This is
the $K_{12}$ set discussed in Sec.~\ref{sec:physics}.

\subsection{Shells of the Barnes--Wall lattice, \texorpdfstring{$d=16$}{d=16}}
\label{app:const_d16}

For $d=16$, we use shells of the Barnes--Wall lattice
$BW_{16}$~\cite{Barnes1959,Nebe2002}. In the normalization used here, its
shortest nonzero vectors have squared norm $4$. The first three nonempty
shells have indices $n=2,3,4$ and yield
\begin{equation}
  m=2160,\quad 30\,720,\quad 261\,360,
\end{equation}
respectively, after retaining one representative from each antipodal pair and
normalizing.

The set used for the candidate dimension witness in Sec.~\ref{sec:physics}
comes from the fifth nonempty shell. It has index $n=6$, squared norm $12$,
and yields $m=4\,480\,320$ unit vectors. The full shell populations are
tabulated in OEIS~A008409~\cite{OEISA008409}.

\subsection{Shells of the Leech lattice, \texorpdfstring{$d=24$}{d=24}}
\label{app:const_d24}

For $d=24$, we use the first two nonempty shells of the Leech lattice
$\Lambda_{24}$~\cite{Conway1999}. They have squared norms $4$ and $6$,
corresponding to shell indices $n=2$ and $n=3$.

We generate these shells using the extended binary Golay code, which specifies
the allowed coordinate patterns and signs~\cite{Conway1999}. In our
integer-coordinate representation, a lattice vector is written as
$w=a/\sqrt{8}$ with $a\in\mathbb{Z}^{24}$. The two shells therefore satisfy
\begin{equation}
  \sum_{j=1}^{24}a_j^2=32
  \quad\text{or}\quad
  \sum_{j=1}^{24}a_j^2=48,
\end{equation}
in addition to the Golay-code constraints.

Retaining one representative from each antipodal pair and setting
$v=a/\|a\|_2$ gives $m=98\,280$ and $m=8\,386\,560$ unit vectors in
$\mathbb{R}^{24}$, respectively. The corresponding shell populations are
tabulated in OEIS~A008408~\cite{OEISA008408}.

\bibliography{arXiv_submission/ref_KG}

\end{document}